\documentclass[12pt,a4paper]{article}

\usepackage[greek,american]{babel}
\usepackage{amssymb}
\usepackage{amsfonts}
\usepackage{amsmath}
\usepackage{relsize}
\usepackage{subfigure}
\usepackage{float}
\usepackage{cite}
\usepackage[pdftex]{graphicx}
\usepackage{dcolumn}
\usepackage{bm}
\usepackage{subfloat}

\usepackage{adjustbox}

\def\P{\mathbb P}
\def\R{\mathbb R}

\def\E{\mathbb E}
\def\X{\mathbb X}
\def\l{\lambda}
\def\a{\alpha}
\def\b{\beta}
\def\ux{\underline x}
\def\ox{\overline x}

\newcommand*{\medcup}{\mathbin{\scalebox{1.5}{\ensuremath{\cup}}}}%

\begin{document}
	
	\begin{center}
	{{{\Large\sf\bf A Bayesian Nonparametric Approach to Reconstruction and Prediction of Random Dynamical Systems}}}\\
	
	\vspace{0.5cm}
	{\large\sf Christos Merkatas, Konstantinos Kaloudis and Spyridon J. Hatjispyros
		\footnote{ Corresponding author. Tel.:+30 22730 82326\\
			\indent email address: schatz@aegean.gr }
	}

	\vspace{0.2cm}
	\end{center}
	\centerline{\sf Department of Mathematics, University of the Aegean}
	\centerline{\sf Karlovassi, Samos, GR-832 00, Greece.} 	
		\date{\today}
		
	\begin{abstract}
		We propose a Bayesian nonparametric mixture model for the reconstruction and prediction  
		from observed time series data, of discretized stochastic dynamical systems, based on Markov Chain Monte Carlo methods (MCMC).
		Our results can be used by researchers in physical modeling interested in a fast and accurate estimation of low
		dimensional stochastic models when the size of the observed time series is small and the noise process (perhaps) is non-Gaussian. The inference procedure is demonstrated
		specifically in the case of polynomial maps of arbitrary degree and when a Geometric Stick Breaking mixture process prior over the space of densities, is
		applied to the additive errors.
		
		Our method is parsimonious compared to Bayesian nonparametric techniques based on Dirichlet process mixtures, flexible and general.
		Simulations based on synthetic time series are presented.
		
		\vspace{0.1in} 
		\end{abstract}

	\section{Introduction}
	\label{sec:Intro}
	
	During the last three decades, nonlinear dynamical systems have been used to explain and model multiple time varying phenomena, exhibiting complex and irregular characteristics \cite{ott2002chaos}, finding applications in different fields of science such as physics, biology, computer science and economics. The erratic and unpredictable behavior of chaotic dynamics was early related to probabilistic and statistical methods of analysis \cite{berliner1992statistics, chatterjee1992chaos}. Nonlinearity alone though, is often not enough to properly describe the evolution of real physical phenomena, so the effect of noise has to be taken into account. In this respect, the constructed predictive model consists of two parts, the nonlinear-deterministic component and the random noise. 
	
	The source of the random noise influencing the procedure of interest is of great importance. If the origin of the noise is the uncertainty of the measurement process, then the available observations can be considered as the corruption of the true system states by measurement-observational noise and the dynamics of the process are not influenced. A widespread approach to confront this type of noise is the application of time delay embedding techniques and related methods, originating from the work of Takens \cite{Ruelle}; see for example the review paper on the analysis of observed chaotic data by Abarbanel \cite{abarbanel2012analysis} and the book of Kantz \cite{kantz2004nonlinear} on the analysis of nonlinear time series and references therein.
	
	Nevertheless, dynamical noise can drastically modify the underlying deterministic dynamics\cite{jaeger1997homoclinic}, as it represents the error in the assumed model, thus compensating for a small number of degrees of freedom. Dynamical systems subjected to the effects of dynamical noise are known as random dynamical systems \cite{Arnold,smith2000nonlinear} and have many applications, mainly because dynamical noise is often present in real data. For the case of dynamical noise, the application of methods based on deterministic inference are not efficient, so many different methods have been proposed regarding the various aspects of the problem. In \cite{muldoon1998delay} a theorem was formulated to cope with the embedding problem for random dynamical systems, requiring multivariate observations. In \cite{siegert1998analysis} and \cite{siefert2004reconstruction} the issue of dynamical reconstruction was addressed for continuous time systems, by estimating drift and diffusion parameters of a Fokker-Plank equation under different types of perturbations. Due to the different impact of the noise types, the goal of discriminating between measurement and dynamical noise, as well as estimating the noise density, is highly significant \cite{heald2000estimation,strumik2008influence,siefert2004differentiate}.
	
	Bayesian formulation \cite{robert2007bayesian} has been of great use in the general field of noise perturbed dynamical systems. It was initially demonstrated in this context by Davies \cite{davies1998nonlinear}, where MCMC methods were used for nonlinear noise reduction. In \cite{Meyer2000} and \cite{Meyer2001} MCMC methods were applied for the parameter estimation of state-space nonlinear models, extending maximum likelihood-based existing methods \cite{McSharry}. Later, in \cite{smelyanskiy2005reconstruction} a path integral representation was proposed for the likelihood function, in order to make inference in stochastic nonlinear dynamics, extended for nonstationary systems in \cite{luchinsky2008inferential}. In \cite{matsumoto2001reconstructions} and \cite{nakada2005bayesian} Bayesian methods were suggested for reconstruction and prediction of nonlinear dynamical systems. Recently in \cite{molkov2012random}, a Bayesian technique was proposed for the prognosis of the qualitative behavior of random dynamical systems under different forms of dynamical noise. 
	
	In this work, we will use a Bayesian approach to reconstruct and predict random dynamical systems. A common assumption in the literature is the normality of the noise process. Such an assumption cannot always be justified and can cause inferential problems when the noise process departs from normality, for example when it produces outlying errors. Then the estimated variance of the normal errors is artificially enlarged causing poor inference for the system parameters of interest. So for example we could have two sources of random perturbations. An environmental source caused by spatiotemporal inhomogeneities \cite{strumik2008influence} producing weak and frequent perturbations, and, a high dimensional deterministic component interpreted in our model as stronger but less frequent perturbations in the form of outlying errors.
	Other cases include systems containing impulsive noise \cite{shinde1974signal,middleton1977statistical}, where the noise probability density function does not decay in the tails like Gaussian. Also, in situations where the system under consideration is coupled to multiple stochastic environments, the driving noise term may exhibit non-Gaussian behavior, see for example references \cite{kanazawa2015minimal} and \cite{kanazawa2015asymptotic}. It is our intention therefore to model the dynamical noise using a highly flexible family of density functions, providing a Bayesian nonparametric formulation \cite{Ferguson,Fuentes}. We are confident that, contrary to the assumption of normality, our Bayesian modeling will be able to capture the right shape of the true underlying noise density hence leading to an improved and reliable statistical inference for the system even in cases where the size of the observed time series is small. Thus, if the number of noise sources is arbitrary, our approach is able to identify the true underlying model. This is  because the different noise sources are represented by the active components of an infinite random mixture. Some recent applications of Bayesian nonparametric methods in nonlinear dynamical systems include Dirichlet process (DP) based reconstruction \cite{Hatjispyros_CSDA_DPM} and joint state-measurement noise density estimation with non-Gaussian and Gaussian observational and dynamical noise components respectively \cite{jaoua2013bayesian}. In this work we aim to:
	\begin{enumerate}
		\item 
		Reconstruct dynamical equations and predict future values, by setting as a prior for the noise process a geometric stick breaking (GSB) mixture \cite{Fuentes} which is effectively a random infinite mixture of probability kernels. Such a reconstruction involves the estimation of the unknown parameters of the deterministic part of the model, the initial condition responsible for the observed noisy time series and density estimation of the unknown and perhaps non-Gaussian error process.
		\item 
		Provide evidence, that modeling discrete time random dynamical systems via GSB mixtures, is efficient, faster and less complicated when compared to Bayesian nonparametric modeling via DP mixtures \cite{Hatjispyros_CSDA_DPM}.
		\item 
		Show that sampling from the posterior joint distribution of the parameters, the initial condition and the future-unobserved-observation variables of the system, provides us with information for the long term behavior of the underlying process in the form of the  quasi-invariant measure of the system.
	\end{enumerate}
	The layout of the paper is as follows. In section 2, we are  giving some preliminary notions in DP mixture priors, and we derive the two competing nonparametric inferential models. The first model is based on DP mixtures, and we develop its randomized-efficient version. It is based on the model that has been used for the reconstruction of random quadratic maps in \cite{Hatjispyros_CSDA_DPM}. It involves two infinite dimensional parameters in the form of random probability weights and locations. The second one, being our main contribution, is simpler and is based on GSB mixtures leading to a faster estimation algorithm as it involves only one infinite dimensional parameter in the form of locations. The GSB based Gibbs sampler is described in detail in section 3. In section 4 we specialize, for simplicity, to dynamical equations with polynomial nonlinearities to an arbitrary degree and we resort to simulation. We use simulated time series produced by a cubic map exhibiting complex dynamical behavior, that is dynamically perturbed by outlying errors of varying intensity. We compare the performance of the proposed GSB based Gibbs sampler against its randomized DP based and plain parametric counterparts in the quality of reconstruction, out-of-sample forecasting and quasi-invariant measure estimation. To demonstrate the need for the nonparametric approach, we also compare the results with those obtained from a parametric Gibbs sampler assuming just Gaussian noise in the inferential procedure. In section 5 we conclude with a summary and future work. 
	Finally, we offer five appendices as a supplementary material. 
	In the supplementary file, appendix A, we provide embedded Gibbs sampling schemes for the various nonstandard densities arising 
	in the implementation of the Gibbs samplers. In then supplementary file, appendix B, we obtain the invariant set of the deterministic 
	part of the cubic map, which we have used for the generation of the synthetic time series for 
	the illustration of our method. Also, in appendix C, we perform a comparison between the GSB sampler and a simple parametric MCMC, that assumes 
	normal dynamical perturbations using noisy logistic time series. Finally, in appendix D we explain the dynamics exhibited by the cubic map used in our numerical experiments.

	\section{Preliminaries and derivation of the inferential models}
	
	A number of approaches have been proposed for system reconstruction. 
	Maximum likelihood based methods treat the unknown parameters like fixed quantities, which maximize the joint conditional 
	distribution of the observed values given the unknown parameters \cite{McSharry, Kostelich, Hammel}. On the other hand 
	Bayesian methods, assume that the parameters themselves are random variables; any prior knowledge can be incorporated together
	with the likelihood function in the form of the joint prior distribution of the parameters. Using Bayes theorem the posterior 
	density, that is the conditional density of the system parameters given the observed time series, can be obtained. Here the most
	crucial step is to sample from the posterior density and thus to recover the marginal posterior density for each system parameter. 
	
	In related work \cite{Hatjispyros_CSDA_DPM}, the assumption of normal errors is being relaxed; the additive independent and identically 
	distributed (IID) dynamical noise is modeled with a family of density functions based on a Bayesian 
	nonparametric model, the DP mixture model \cite{Lo}. In their approach, they have modeled the random noise to have density
	\begin{equation} 
	\label{dpmodel} 
	f_{\mathbb P}(z) = \int_{v>0} {\cal N}\left(z|\,0, v^{-1}\right)\P\,(dv),
	\end{equation}
	where $v$ is the precision of the zero mean normal distribution ${\cal N}\left(z|\,0, v^{-1}\right)$ and 
	$\P=\sum_{j\ge 1}w_j\delta_{\l_j}\!\sim\!{\cal DP}(c,{\rm P}_{0})$, 
	is a discrete random probability measure defined over ${\mathbb{R}^{+}}$, drawn from a DP \cite{Ferguson}
	with concentration parameter $c>0$ and base measure ${\rm P}_0(dv)={\cal G}(v|a,b)dv$, a gamma measure with shape $a$ and rate $b$. 
	The Dirac measures $\delta_{\l_j}$ are concentrated on the random precisions $\l_j$ (the locations of $\P$) 
	which are IID from ${\rm P}_0$. 
	The random probability weights $w_i$ are defined via a stick-breaking process \cite{Sethuraman82, Sethuraman94} so that 
	$w_1=z_1$ and for $j>1$
	\begin{equation}
	\label{stickbreakingw}
	w_j = z_j\prod_{s<j} (1-z_s),
	\end{equation}
	with the $z_i$ variables IID from ${\cal B}e(1,c)$, a beta distribution with mean $(1+c)^{-1}$.
	
	Our intention is to reconstruct dynamical equations and jointly predict future values, by modeling additive noise components as
	geometric stick breaking mixture (GSBM) processes. We will show that GSBM modeling is as accurate as DPM modeling but less complicated and faster. 
	Effectively, we will substitute the mixing measure ${\mathbb P}$  in equation (\ref{dpmodel}) 
	by the random measure $\P_{\rm G}$ given by  
	$$
	\P_{\rm G}=p\sum_{j=1}^\infty \,(1-p)^{j-1}\delta_{\lambda_{j}}
	\,\,\,{\rm with}\,\,\,p=(1+c)^{-1},
	$$
	first introduced in Fuentes--Garcia et al. (2010). Mind that, although the random measure $\P_{\rm G}$ is closely related to $\P\sim{\cal DP}(c,{\rm P}_0)$
	(the measure $\P_{\rm G}$ is the expectation of $\P$ given the random locations $\l_j$), $\P_{\rm G}$ is not of the Dirichlet type and is not a conjugate prior
	over the space of measures
	(see Ref. \cite{Mena} and references therein).

	\subsection{The Model}
	
	We consider the following random dynamical model given by
	\begin{equation} \label{model}
	X_{i} = T(\theta, X_{i-1}, Z_{i})=g(\theta, X_{i-1}) + Z_{i},\quad i\geq 1,
	\end{equation}
	where $g:\Theta\times {\mathbb X}\to {\mathbb X}$, for some compact subset ${\mathbb X}$ of $\mathbb R$,
	$(X_i)_{i\ge 0}$ and $(Z_i)_{i\ge 1}$\ are real random variables over some probability space $(\Omega, {\cal F}, {\rm P})$; 
	the set $\Theta$ denotes the parameter space and $g$ is nonlinear, and for simplicity, continuous in $X_{i-1}$. 
	We assume that the random variables $Z_i$ are independent to each other, and independent of the states $X_i$. 
	In addition we assume that the additive perturbations $Z_i$ are identically distributed from a zero mean 
	distribution with unknown density $f$ defined over the real line, so that $T:\Theta\times {\mathbb X}\times{\mathbb R}\to {\mathbb R}$.
	We assume that there is no observational noise, so that we have at our disposal a time series $X^n=(x_1,\ldots, x_n)$ generated by 
	the Markovian processes defined in equation (\ref{model}). The time series $X^n$ depends solely on the initial distribution of $X_0$, 
	the vector of parameters $\theta$, and the particular realization of the noise process. 
	
	We will model the errors in recurrence relation (\ref{model}) as a random infinite mixture of zero mean normal kernels.
	As a mixing measure, initially we will use a general discrete random distribution 
	${\mathbb G}=\sum_{j\ge 1}\pi_j\,\delta_{\lambda_j}$ with random probability weights $\pi=(\pi_j)_{j\ge 1}$ and locations $\lambda=(\lambda_j)_{j\ge 1}$; then
	the conditional density of $x$ given $\pi$ and $\lambda$ can be represented as 
	$$
	f_{\pi,\lambda}(x)=\int_{\R^+}{\cal N}\left(x|\,0,v^{-1}\right){\mathbb G}(dv)
	=\sum_{j=1}^\infty\pi_j\, {\cal N}\left(x|\,0,\lambda_j^{-1}\right).
	$$
	Conditionally on the variable $x_0$, we have the transition kernels 
	\begin{equation}
	\label{generalSBmeasure}
	f_{\pi,\lambda}(x_i|\,x_{i-1},\theta)=\sum_{j=1}^\infty
	\pi_j\, {\cal N}\left(x_i|\,g(\theta,x_{i-1}),\lambda_j^{-1}\right),\,\,1\le i\le n,
	\end{equation}
	with associated data likelihood
	$$
	f_{\pi,\lambda}\left(x_1,\ldots,x_n|\,x_0,\theta\right)=\prod_{i=1}^n\sum_{j=1}^\infty\pi_j\, {\cal N}\left(x_i|\,g(\theta,x_{i-1}),\lambda_j^{-1}\right). 
	$$

	\subsection{Dynamical Slice Sets}
	Due to the infinite mixture appearing in the product of the likelihood in the equation above, we are not able to construct Gibbs samplers of finite dimensions. To make the number of variables that we have to sample finite, we use slice sampling techniques for infinite mixtures.
	For each observation $x_i$, we introduce  the pair $(d_i, A_i)$ where $d_i$ is the random variable
	that indicates the component of the infinite mixture the observation $x_i$ came from, and $A_i$ the associated $x_i$-slice set, which
	is a random almost surely finite set of indices.
	Marginally, we select each $d_i$ with probability $\pi_i$ that is
	$(d_i|\,\pi)\sim\sum_{j\ge 1}\pi_j\,\delta_j$, 
	and the random variables $d_i$ have an infinite state space. 
	To have $(x_i|\lambda,A_i)$ coming from a finite mixture of normal kernels, a prerequisite that will enable us to create a Gibbs sampler with a finite number of updates, we let 
	$d_i$ conditionally on the event $(d_i\in A_i)$, to attain a discrete uniform distribution, that is
	\begin{eqnarray}
	f_\lambda(x_i|\,A_i)  & = &  \sum_{j=1}^\infty f_\lambda(d_i=j|\,A_i)\,f_\lambda(x_i|\,d_i=j)\nonumber\\
	&=&  \sum_{j\in A_i}|A_i|^{-1}{\cal N}\left(x_i|\,0,\lambda_j^{-1}\right),\nonumber
	\end{eqnarray}
	where $|A_i|$ denotes the cardinality of the set $A_i$.
	Thus, given the precisions $\lambda$ and the set $A_i$, the observation $x_i$ comes from an equally weighted and almost surely finite mixture of normals.
	
	We will consider two types of slice sets:
	
	\vskip0.05in\noindent{\bf Non sequential slice sets.}
	To each observation $x_i$, we assign the set $A_i=\{j\in{\mathbb N}:0<u_i<\pi_j\}$ \cite{Walker}
	that depends on the weights $\pi$ through the random variable $u_i$ such that 
	$f_{\pi}(d_i=j|\,u_i)\propto\pi_j\,{\cal U}(u_i|\,0,\pi_j)$,
	where ${\cal U}(x|\,0,\pi_j)$ denotes the uniform density over the interval $(0,\pi_j)$. Letting $\pi=w$, with $w$
	the stochastically ordered probability weights introduced in (\ref{stickbreakingw}), for $1\le i\le n$ we obtain a DP mixture based augmented random density
	\begin{eqnarray}
	\label{gapsmodel}
	f_{w,\lambda}(x_i,u_i, d_i=j) &=&w_j\,{\cal U}(u_i|\,0,w_j)\nonumber\\
	& &\times \,{\cal N}(x_i|\,0,\lambda_j^{-1}).
	\end{eqnarray}
	
	\vskip0.05in\noindent{\bf Sequential slice sets.}
	Letting $A_i=\{1,\ldots,N_i\}$, with $N_i$ being an almost surely finite discrete random 
	variable of mass $f_N(\,\cdot\,|\,p)$, and letting 
	$f(d_i=j|\,N_i)=N_i^{-1}\,{\cal I}(j\le N_i)$ for $1\le i\le n$ a discrete uniform distribution on the set $A_i$, we obtain a GSB mixture based augmented random density 
	\begin{eqnarray}
	\label{nogapsmodel}
	f_{\lambda}(x_i,N_i=l, d_i=j) &=&\,f_N(l\,|\,p)\,l^{-1}\nonumber\\
	& & \times\,{\cal I}(j\le l)\,{\cal N}(x_i|\,0,\lambda_j^{-1}).
	\end{eqnarray}
	Marginalizing (\ref{nogapsmodel}) with respect to $(N_i,d_i)$, it is that
	$$
	f_{\lambda}(x_i)\,=\,\sum_{j=1}^\infty\pi_j\,{\cal N}(x_i|\,0,\lambda_j^{-1})
	\,\,\,{\rm with}\,\,\,\pi_j=\sum_{l=j}^\infty l^{-1}f_N(l\,|\,p).
	$$
	When $N_i$ comes from the negative binomial
	distribution $f_N(l\,|\,p)={\cal N}\!{\cal B}(l\,|\,2,p)=lp^2(1-p)^{l-1}{\cal I}(l\ge 1)$, the weights $\pi_j$ are geometric, that is 
	\begin{equation}
	\label{geometricweights}
	\pi_j =\sum_{l=j}^{\infty}l^{-1}f_{N}(l\,|\,p) = \sum_{l=j}^{\infty}l^{-1}lp^2(1-p)^{l-1} = p\,(1-p)^{j-1}.
	\end{equation}
	The geometric weights can be thought of as a reparametrization of the expectation of the stick-breaking weights given in 
	(\ref{stickbreakingw}), in the sense that $\pi_j=\E(w_j)$ with $p=(1+c)^{-1}$.  Other  distributions could be used as well but it is the ${\cal NB}(l\,|\,2,p)$ that leads to a model with lesser complexity. It can be shown that the use of a ${\cal NB}(l\,|\,k,p)$ will give to the weights the form of a equally weighted mixture of Negative Binomial distributions that is $\pi_j=(k-1)^{-1}\sum_{r=1}^{k-1}{\cal NB}(l\,|\,r,p).$
	
	\subsection{The two dynamical reconstruction models}
	
	Now it becomes clear, that depending on the choice of the slice sets, we obtain two types of dynamical reconstruction models.
	
	\vskip0.05in\noindent{\bf 1. The DP mixture based model:}
	From relations (\ref{generalSBmeasure}) and (\ref{gapsmodel}) it is that 
	\begin{eqnarray}\label{partialdpdatalike}
	f_{w,\lambda}(x_i, u_i, d_i=j|\,x_{i-1},\theta) &=& w_j\,{\cal U}(u_i|\,0,w_j)\nonumber\\
	& \times& {\cal N}\left(x_i|\,g(\theta,x_{i-1}), \lambda_j^{-1}\right).
	\end{eqnarray}
	
	
	\noindent 
	In a hierarchical fashion using the slice variables $u_i$ and the stick-breaking representation we have that
	for $i=1,\ldots,n$ and $j\ge 1$:
	\begin{align}
	\nonumber
	&  (x_i|\, x_{i-1}, d_i=j, \theta, \lambda)\,\stackrel{\rm IND}\sim\,{\cal N}(g(\theta,x_{i-1}), \lambda_j^{-1})\\
	\nonumber
	&  (u_i|\,d_i=j,w)\,\stackrel{\rm IND}\sim\, {\cal U}(0,w_j)\\
	\nonumber
	&  {\rm P}(d_i=j|\,w)=w_j\nonumber\\
	\nonumber
	&   w_j=z_j\mathop{\mathsmaller{\prod}}\nolimits_{s<j}(1-z_s),\,\,z_j \,\stackrel{\rm IID}\sim{\cal B}e(1,c)\\
	\nonumber
	&   \lambda_j\,\stackrel{\rm IID}\sim\,P_0.
	\end{align}
	
	\smallskip\noindent{\bf 2. The GSB mixture based model:}
	From relations (\ref{generalSBmeasure}) and (\ref{nogapsmodel}), and letting $f_N(\,\cdot\,|p)={\cal N}\!{\cal B}(\,\cdot\,|2,p)$, we have
	\begin{eqnarray}\label{partialgsbdatalike}
	f_{\lambda}(x_i, N_i=l, d_i=j|\,x_{i-1},\theta) &=&{\cal N}\!{\cal B}(l|\,2,p)\,l^{-1}\,{\cal I}(j\le l)\nonumber\\
	&\times&{\cal N}\left(x_i|\,g(\theta,x_{i-1}), \lambda_j^{-1}\right).\nonumber\\
	\end{eqnarray}
	
	\noindent
	In a hierarchical fashion using the slice variables $N_i$ we have that
	for $i=1,\ldots,n$ and $j\ge 1$:
	\begin{eqnarray}
	& & (x_i|\, x_{i-1}, d_i=j, \theta, \lambda)\,\stackrel{\rm IND}\sim\,
	{\cal N}(g(\theta,x_{i-1}), \lambda_j^{-1})\nonumber\\
	& & (d_i|\,N_i=l)\,\stackrel{\rm IND}\sim\, {\cal U}\{1,\ldots,l\}\nonumber\\
	& & \pi_j={\cal N}\!{\cal B}(j|\,1,p),\,\,N_i\,\stackrel{\rm IID}\sim\,{\cal N}\!{\cal B}(2,p)\nonumber\\
	& & \lambda_j\,\stackrel{\rm IID}\sim\,P_0,\nonumber 
	\end{eqnarray}
	
	We have the following proposition:
	
	\smallskip\noindent
	{\bf Proposition $1$.} {\sl 
		The likelihoods $f_{w,\lambda}^{\cal D}$ and $f_\lambda^{\cal G}$ that include the $T$ future unobserved values $(x_{n+1},\ldots,x_{n+T})$ of the observed series $X^n$, 
		based on the DPM and GSBM models respectively, are given  by:

			\begin{equation}\label{dpdatalike}
			f_{w,\lambda}^{\cal D}(x_i, u_i, d_i,i=1,\ldots, n_T|\,\theta,x_0,c)\propto
			\prod_{1\le i\le n_T\atop d_i;\, u_i<w_{d_i}}\sqrt{\lambda_{d_i}}\exp\left\{-{\lambda_{d_i}\over 2}h_\theta(x_i,x_{i-1})\right\}, 
			\end{equation}
			and
			\begin{equation}\label{gsbdatalike}
			f_\lambda^{\cal G}(x_i, N_i, d_i,i=1,\ldots, n_T|\,\theta, x_0, p)\propto
			\prod_{1\le i\le n_T\atop d_i;\,d_i\le N_i} p^2(1-p)^{N_i-1}\sqrt{\lambda_{d_i}}\exp\left\{-{\lambda_{d_i}\over 2}h_\theta(x_i,x_{i-1})\right\},
			\end{equation}

		where $h_\theta(x_i,x_{i-1})=(x_i-g(\theta,x_{i-1}))^2$ and $n_T=n+T$. }
	
	\smallskip\noindent
	{\bf Proof.} The expressions for the two augmented data-likelihoods $f_{w,\lambda}^{\cal D}$ and $f_\lambda^{\cal G}$ are coming from equations 
	(\ref{partialdpdatalike}) and (\ref{partialgsbdatalike}) and their corresponding hierarchical representations. \hfill$\square$
	
	\smallskip\noindent
	It is now clear from the form of the likelihood that a Gibbs sampling scheme will have finite number of updates.
	The details of the GSB mixture based reconstruction model (from now on referred to as the GSBR model) is now described in Section $3$. 
	The implementation of the algorithm and further details involving the DP mixture based model, here generalized to a random concentration mass $c$ (from now on referred to as the rDPR model) can be found in Hatjispyros et al. (2009).

	\section{Sampling algorithms}
	
	To choose the fittest between the rDPR and GSBR models, we adapt to a ``synchronized'' prior specification.
	More specifically, in this paper we use a fully stochastic version of the DPR algorithm,  
	which involves imposing a ${\cal G}(\alpha,\beta)$ prior over the concentration parameter $c$ as proposed by West in Ref \cite{west1992hyperparameter}.
	(we remark that in \cite{Hatjispyros_CSDA_DPM}, the concentration parameter $c$  has been set to $c=1$ through out the numerical experiments).
	Then, ``synchronized'' prior specifications involve a transformed gamma prior over the geometric probability $p$ via $p=(1+c)^{-1}$.
	So as a prior over $p$ we set
	\begin{equation}
	\label{prioroverp}
	f(p)={\cal TG}(p\,|\,\a,\b)={\b^\a e^\b\over \Gamma(\a)}\,
	p^{-(\alpha+1)}e^{-\beta/p}(1-p)^{\alpha-1},  
	\end{equation} 
	with $p\in (0, 1).$
	Note that for generic applications of the GSBR model, a beta conjugate prior $f(p;\a,\b)={\cal B}e(p;\a,\b)$ is preferable as it leads to an implementation of lesser complexity. We have noticed that both priors provide results that are nearly indistinguishable.
	As a base measure for both models, we use $P_0(d\lambda_j)={\cal G}(\lambda_j|a,b)d\lambda_j,\,j\ge 1$ for fixed hyperparameters $a$ and $b$. 
	
	Having completed the model, we are now ready to describe the Gibbs sampler and the full
	conditional densities for estimating the GSBR model. After initializing the variables $d_i$ for $i=1,\ldots,n_T$ 
	and the variables $p,x_0$ and $\theta$, at each iteration, we will sample the variables:
	$$
	(\lambda_j),\,1\le j\le N^*,\quad (d_i, N_i),\,1\le i\le n_T,
	$$
	and
	$$
	(\theta,\,x_0,\,p,\,z_{n_T+1}),
	$$
	with $N^*=\max_{1\le i\le n_T}N_i$.
	
	\medskip\noindent {\bf 1.}
	We first sample the precisions $\lambda_j$ for $j=1,\ldots,d^*$ and $d^*=\max_{1\le i\le n_T}d_i$. We have that

		\begin{equation*}
		f(\lambda_j\,|\cdots)={\cal G}\left(\lambda_j\,|
		a + {1\over 2}\sum_{i=1}^{n_T}{\cal I}(d_i=j),\, b + {1\over 2}\sum_{i=1}^{n_T}{\cal I}(d_i=j)\,h_\theta(x_i, x_{i-1})\right),
		\end{equation*}

	where the expression $f(\lambda_j\,|\cdots)$ denotes the density of $\lambda_j$
	conditional on the rest of the variables.
	If $N^*>d^*$ we sample the additional $\lambda_j$'s from the prior ${\cal G}(a,b)$.
	
	\medskip\noindent {\bf 2.}
	We then sample the infinite mixture allocation variables $d_i$ for $i=1,\ldots,n_T$. It is that
	$$
	{\rm P}(d_i=j\,|\cdots)\,\propto\, \lambda_j^{1/2}
	\exp\left\{-{\lambda_j\over 2}h_\theta(x_i, x_{i-1})\right\}{\cal I}(j\le N_i).
	$$
	
	\medskip\noindent {\bf 3.}
	Next, to construct the sequential slice sets $A_i$ for $1\le i\le n_T$ 
	we have to sample $N_i$ from
	$$
	{\rm P}(N_i=l\,|\,d_i=j,\cdots)\,\propto\, (1-p)^l\,{\cal I}(l\ge j),
	$$
	which is a truncated geometric distribution over the set $\{j,j+1,\ldots\}$.
	
	\medskip\noindent {\bf 4.} 
	The full conditional for $x_0$, with a uniform prior over the set $\tilde{\X}\subseteq{\mathbb R}$ that 
	represents our prior knowledge for the state space of the dynamical system in relation (\ref{model}) will be
	\begin{equation}
	\label{fullcondx0}
	f(x_0|\cdots)\,\propto\,{\cal I}(x_0\in\tilde{\X})\exp\left\{-{\lambda_{d_1}\over 2} h_\theta(x_1, x_0)\right\}.
	\end{equation}
	
	\medskip\noindent {\bf 5.} 
	The full conditional densities for the future unobserved observations, when $T\ge 2$ and for $j=1,\ldots, T-1$, are given by
		\begin{equation}
		\label{firstTminus1}
		f(x_{n+j}|\cdots)\,\propto\,\exp\left\{-{1\over 2}\left[\lambda_{d_{n+j}}h_\theta(x_{n+j},x_{n+j-1})
		+\lambda_{d_{n+j+1}}h_\theta(x_{n+j+1},x_{n+j})\right]\right\}.
		\end{equation}

	For $j=T$ the full conditional is normal with mean $g(\theta, x_{n+T-1})$
	and variance $\lambda_{d_{n+T}}^{-1}$, that is
	\begin{equation}
	\label{firstT}
	f(x_{n+T}|\cdots)\,=\,{\cal N}\left(x_{n+T}|\,g(\theta, x_{n+T-1}),\lambda_{d_{n+T}}^{-1}\right).
	\end{equation}
	
	\medskip\noindent {\bf 6.}
	For the vector of parameters $\theta$, and assuming a uniform prior over the subset $\tilde{\Theta}$ of the parameter space ${\mathbb R}^k$, the full conditional becomes
	\begin{equation}
	\label{fullcondtheta}
	f(\theta\,|\cdots)\,\propto\,{\cal I}(\theta\in\tilde{\Theta})
	\exp\left\{-{1\over 2}\sum_{i=1}^{n_T}\lambda_{d_i}h_\theta(x_i,x_{i-1})\right\}.
	\end{equation} 
	
	\medskip\noindent {\bf 7.} 
	Taking into consideration relation (\ref{prioroverp}), the full conditional for the geometric probability $p$ is
	\begin{equation}
	\label{pnonstandard}
	f(p\,|\cdots)\,\propto\,p^{2n_T-\alpha-1}\,(1-p)^{L_{n_T}}\,e^{-\beta/p}\,{\cal I}(0<p<1),
	\end{equation} 
	where $L_{n_T}=\alpha+\sum_{i=1}^{n_T} N_i-n_T-1$.
	
	\medskip\noindent {\bf 8.} 
	Having updated $p$, we construct the geometric weights $\pi_j$ for $1\le j\le N^*$ via equation 
	(\ref{geometricweights}). We are now ready to sample $z_{n+1}$ from the noise predictive 
	$f(z_{n+1}|x_1,\ldots,x_n)$.
	At each iteration of the Gibbs sampler we have updated weights $(\pi_j)_{1\le j\le N^*}$ and precisions $(\lambda_j)_{1\le j\le N^*}$
	and we sample independently $\rho\sim{\cal U}(0,1)$. Then we take the $\lambda_j$ with $1\le j\le N^*$ satisfying
	$$
	\sum_{i=0}^{j-1}\pi_i<\rho\le\sum_{i=0}^j\pi_i,\,\,\,\pi_0=0.
	$$
	If $\rho>\sum_{i=0}^{N^*}\pi_i$, we sample $\lambda_j$ from the prior ${\cal G}(a,b)$. In any case we sample $z_{n+1}$ from the 
	normal kernel ${\cal N}(0,\lambda_j^{-1})$.
	
	Details on sampling efficiently via embedded Gibbs samplers, thus circumventing Metropolis-within-Gibbs implementations for the nonstandard densities arising in equations (\ref{fullcondx0}) through
	(\ref{pnonstandard}), are provided in the supplementary file appendix A.

	
	\section{Simulation Results}
	
	Quadratic polynomial maps, can exhibit for each parameter 
	value at most one stable attractor. Multistability and coexistence of more than one strange attractors can
	be achieved under higher degree polynomial maps \cite{nishimotobifurcation,kraut1999preference}. We will generate observations from a cubic random map with a deterministic part given by
	\begin{equation}\label{amap}
	\tilde{g}(\vartheta,x)=0.05+\vartheta x-0.99 x^3.
	\end{equation}
	When $\vartheta\in [\,\underline{\vartheta},\,\overline{\vartheta}\,]$ with $\underline{\vartheta}=-0.04$ 
	and $\overline{\vartheta}=2.81$ the dynamics of $\tilde{g}$, starting from $x_0=1$, are bounded. The map becomes bistable in the regions under the extrema of (\ref{amap}) 
	when  $\vartheta\in \Theta_{\rm bi}=[\,\underline{\vartheta}_{\,\rm bi},\overline{\vartheta}_{\rm bi}]$ with  $\underline{\vartheta}_{\,\rm bi}=1.27$ 
	and $\overline{\vartheta}_{\rm bi}=2.54$.
	When $\vartheta>\overline{\vartheta}_{\rm bi}$, the two coexisting chaotic attractors collapse into one global attractor and the dynamics oscillate between the domains previously occupied by the isolated attractors. For more details concerning the dynamical behavior of the map given in relation (\ref{amap}) we refer to appendix D.
	
	\medskip\noindent {\bf Noise processes:} 
	We illustrate the GSBR and rDPR models with simulated data sets,
	consisting of observations generated from the cubic random recurrence $x_i=\tilde{g}(\vartheta,x_{i-1})+z_i$,
	for the specific parameter value $\vartheta^*=2.55$ and initial condition $x_0=1$.
	The dynamical noise $z_i$ was sampled from:
	
	\noindent
	1. The equally weighted normal 4-mixture
	\begin{equation}
	\label{noisedensity} 
	f_1=\sum_{r=0}^3{1\over 4}{\cal N}\left(0,(5r+1)\sigma^2\right),\,\, \sigma=10^{-2}.
	\end{equation}
	
	\noindent
	2.   The normal 2-mixtures, which exhibit progressively heavier tails for $1\le l\le 4$
	\begin{equation}
	\label{noisedensityout}
	f_{2,l}={5+l\over 10}{\cal N}(0,\sigma^2)+{5-l\over 10}{\cal N}\left(0,(200\sigma)^2\right),\,\,\sigma=10^{-3}. 
	\end{equation}
	
	As a measure of the tail fatness of the density $z\sim f$, we use the mean absolute deviation 
	from the mean normalized by the standard deviation, for a zero mean $z$ it is that
	$TF_f=\E|z|/\sqrt{\E|z|^2}$ . The closer $TF_f$ is to 1, the thinner the tails are. It can be verified numerically that
	$$
	TF_{f_1}>TF_{f_{2,1}}>\cdots>TF_{f_{2,4}}.
	$$
	We model the deterministic part $g(\theta,x)$ of the map in equation (\ref{model}) with a polynomial in $x$ of degree $m=5$.
	
	Our findings is that the GSBR models are more amenable to dynamical reconstruction 
	purposes; they are as accurate as the DPR models, they give smaller execution times and are less complicated and thus easier to implement. In all the examples we also compare the results with the results obtained from a parametric reconstruction and prediction Gibbs sampler, that is assuming just Gaussian noise. We refer to this model as Param in the tables.

	\medskip\noindent {\bf Prior specifications:} Here we define the synchronized prior specifications of the GSBR and rDPR Gibbs samplers. We use the following general prior set up:
	\begin{align}
	\nonumber 
	& c\sim{\cal G}(\a,\b),\quad p\sim{\cal TG}(\a,\b),\quad\{\lambda_j\sim{\cal G}(a,b):\, j\ge 1\}\\
	\nonumber 
	& \theta\sim{\cal U}((-M,M)^{k+1}),\quad x_0\sim{\cal U}(-M_0,M_0), 
	\end{align}
	where $k$ is the degree of the modeling polynomial.
	
	\noindent 
	{\bf A. Noninformative reconstruction and prediction:}
	In the absence of any prior knowledge, we propose
	a noninformative prior specification for simultaneous reconstruction and prediction, namely
	$$
	{\cal PS}_{\rm NRP}:\,\a=\b\ge 10^{-1},\,\,  a=b\ge 10^{-4},\, M\gg 1,\, M_0\gg 1. 
	$$
	{\bf B. Informative reconstruction and prediction:}
	When a-priori we believe that the dynamical noise resembles a finite mixture of zero mean Gaussians  
	with variances that are close to each other, we set:
	$$
	{\cal PS}_{\rm IRP}:\,\a>\b\ge 10^{-1},\,\, a>b\ge 10^{-4},\, M\gg 1,\, M_0\gg 1. 
	$$ 
	Such prior specifications induce a small average GSB probability $p$  
	(and consequently a large average DP concentration mass $c$), 
	forcing the Gibbs samplers to activate a large number of normal kernels. Thus generating
	a more detailed Gaussian mixture representation of the unknown dynamical noise.
	
\begin{figure*}
		\centering
		\includegraphics[width = 1\textwidth]{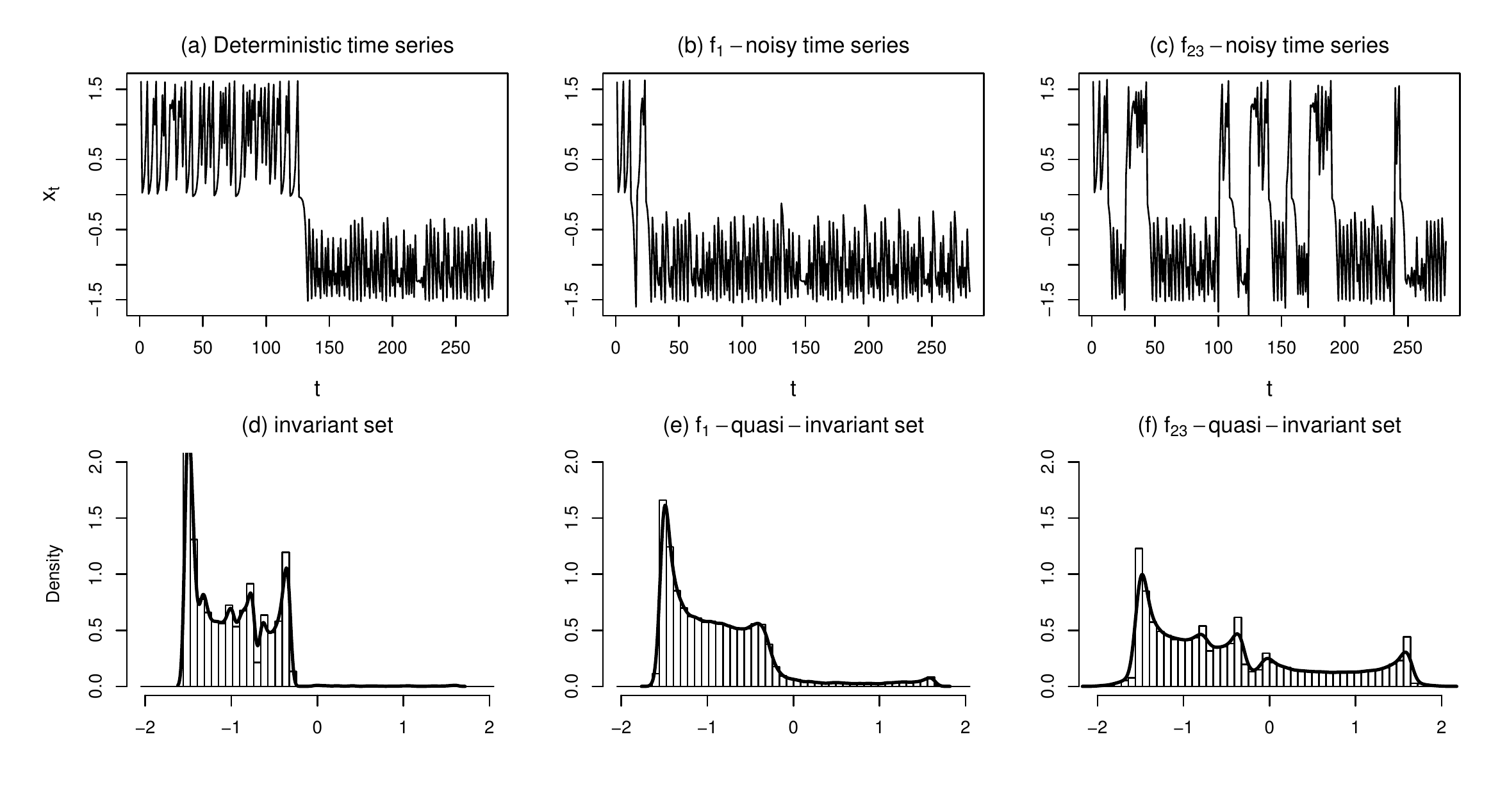}
		\caption{In figures \ref{fig1}(a)-(c) we display the deterministic orbit and $f_1$ and $f_{2,3}$ data-realizations with initial condition $x_0=1$. 
			In figures \ref{fig1}(d)-(f) we display the deterministic invariant density approximation and the $f_1$ and $f_{2,3}$ quasi-invariant densities approximations respectively.
			\label{fig1}}
	\end{figure*}

	\medskip
	\noindent 
	{\bf Data sets and invariant sets:}  
	In Figure \ref{fig1}(a), we display the deterministic orbit of length $280$ of the deterministic map 
	$y_i=\tilde{g}(\vartheta^*,y_{i-1})$, with starting point at $y_0=1$. We have approximated the interval $\mathbb X$ that is remaining invariant 
	under the action of $\tilde{g}(\vartheta^*,\,\cdot\,)$ by $[-1.8881,1,8991]$ (see supplementary file, Appendix B), and the associated average characteristic Liapunov exponent by $0.4625$. 
	Realizations of the random recurrence 
	$x_i=\tilde{g}(\vartheta^*,x_{i-1})+z_i,\,\,x_0=1$ under different types of noise are given  in Figures \ref{fig1}(b) and \ref{fig1}(c) respectively. 
	
	Our observations for reconstruction and out-of-sample prediction will be the data sets $X_{f_1}^{200}$ and 
	$\{X_{f_{2,l}}^{200}:1\le l\le 4\}$. The latter data sets, have been generated in R under the random number generator seeds $RNG_{f_1}=1$ and $RNG_{f_{2,l}:1\le l\le 4}=\{10,15,13,38\}$.
	Approximations of the deterministic and noisy invariant measures are given in Figures \ref{fig1}(d)-(f).  
	The deterministic invariant measure $\mu_{{\tilde g},0}(dy)$ is approximated
	in Figure \ref{fig1}(d). 
	The $z$-noisy measures $\mu_{{\tilde g},z}(dx)$ approximated in
	Figures \ref{fig1}(e) and \ref{fig1}(f), are quasi-invariant in the sense that for all measurable subsets $B$ of $\mathbb R$ it is that
	$\mu_{{\tilde g},z}(B)=\lim_{t\to\infty}{\rm P}(x_t\in B\,|\,\tau_{{\mathbb X}'}>t)$,
	where $\tau_{{\mathbb X}'}$ is a random time denoting the first time the system enters the trapping set ${\mathbb X}'$ (see supplementary file, appendix B).
	
	\medskip
	\noindent{\bf Complexity measures and prior specifications:} 
	The occurrence of an informative structure in the available data sets,  
	may help the practitioner to decide between an informative and a noninformative prior set up.
	Approximate entropy (ApEn) 
	\cite{Pincus, BorchersPracma} can be used to assess the complexity of the available set $x_f^{(n)}$ of observations. Large ApEn values, indicate irregular and unpredictable time series data. 
	Nevertheless, it is known that ApEn values are heavily dependent on sample size (lower than expected for small sample sizes).
	A recently developed complexity measure that is  less dependent on the sample size, is the forecastable component analysis $\Omega$ (ForeCA) \cite{Goerg, GoergForeCA}, which is based on the 
	entropy of the spectral density of the time series, and is normalized between zero and one.  Large $\Omega$ values characterize more predictable time series.
	
\begin{figure}
		\centering
		\includegraphics[width = 1\textwidth]{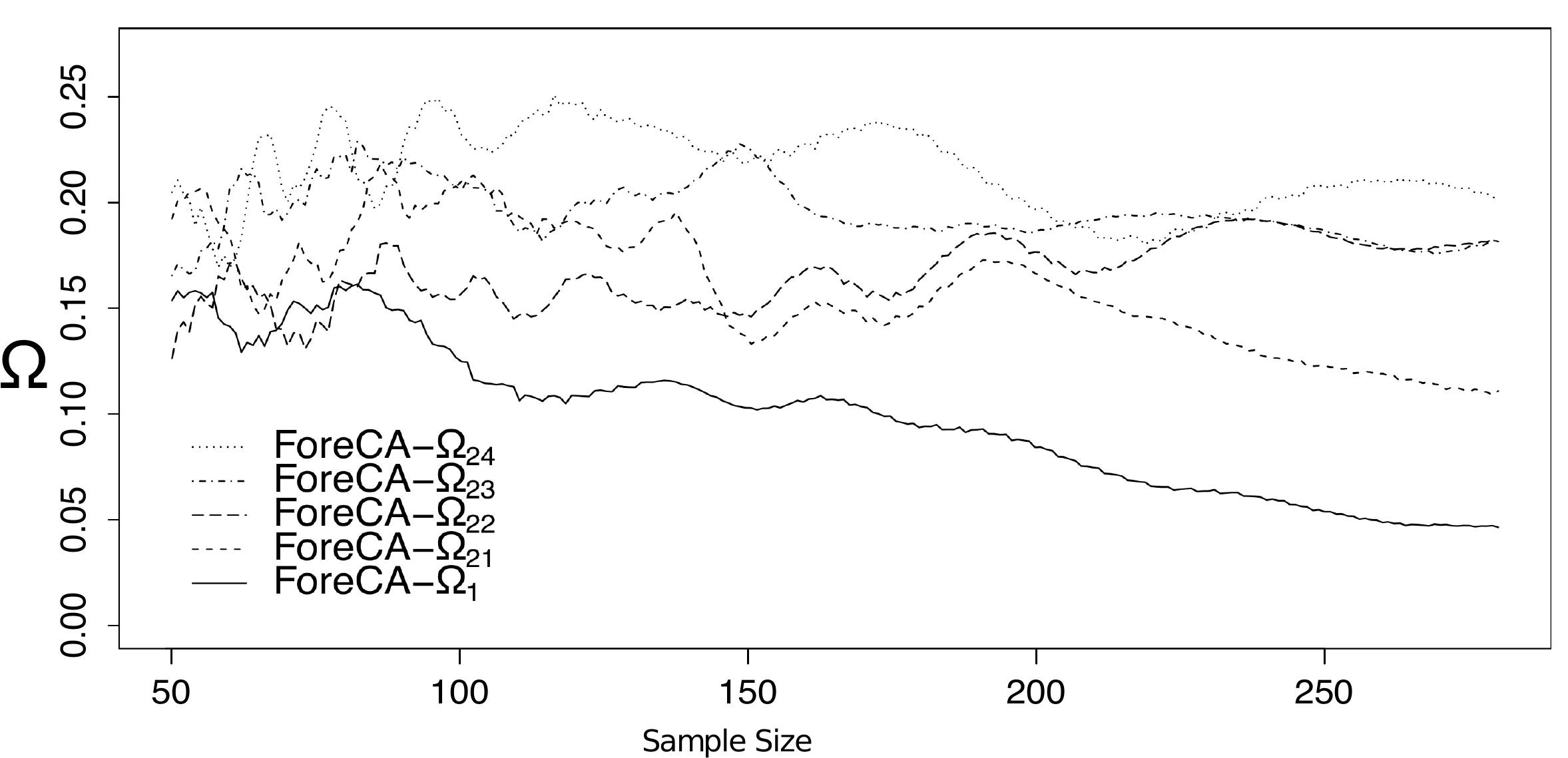}
		\caption{Here we display the $\Omega$ curves relating to the data sets $X_{f_1}^n$ and  $\{X_{f_{2,l}}^n:1\le l\le 4\}$ for $n$ between 50 and 280.
			\label{fig2}}            
	\end{figure}

	In Figure \ref{fig2} we display the $\Omega$ curves as functions of the sample size $n$, for the time series $X_{f_1}^n$ and $\{X_{f_{2,l}}^n:1\le l\le 4\}$. For the computation of the $\Omega$ curves we have used the weighted overlapping segment averaging (WOSA) method \cite{GoergForeCA}.
	The data sets  $\{X_{f_{2,l}}^n:1\le l\le 4\}$ have the more informative structure as for $n>80$ and $1\le l\le 4$
	it is that 
	$$
	\Omega(X_{f_{2l}}^n)>\Omega(X_{f_1}^n).
	$$

	\subsection{Informative reconstruction and prediction under the $f_1$ dynamic noise} 
	
	We ran the Param, rDPR and GSBR Gibbs samplers for $T=20$ in a synchronized mode, for $5\times 10^5$ iterations and a burn-in period of $10,000$,
	using data set $X_{f_1}^{200}$ under the informative prior specification ${\cal PS}_{\rm IRP}$ with $\a=3,\,\b=0.3$, $a=1,\,b=10^{-3}$ and $M=M_0=10$. We remark that under noninformative prior specifications of the form $\a=\b\le 0.3$, and $a=b\le 10^{-3}$,
	the average number of active normals for  both nonparametric samplers
	is lesser than four, leading to less accurate estimations. The following provide a summary and some brief comments.
	
\begin{figure}
		\centering
		\includegraphics[width=1\textwidth,height=0.5\textheight]{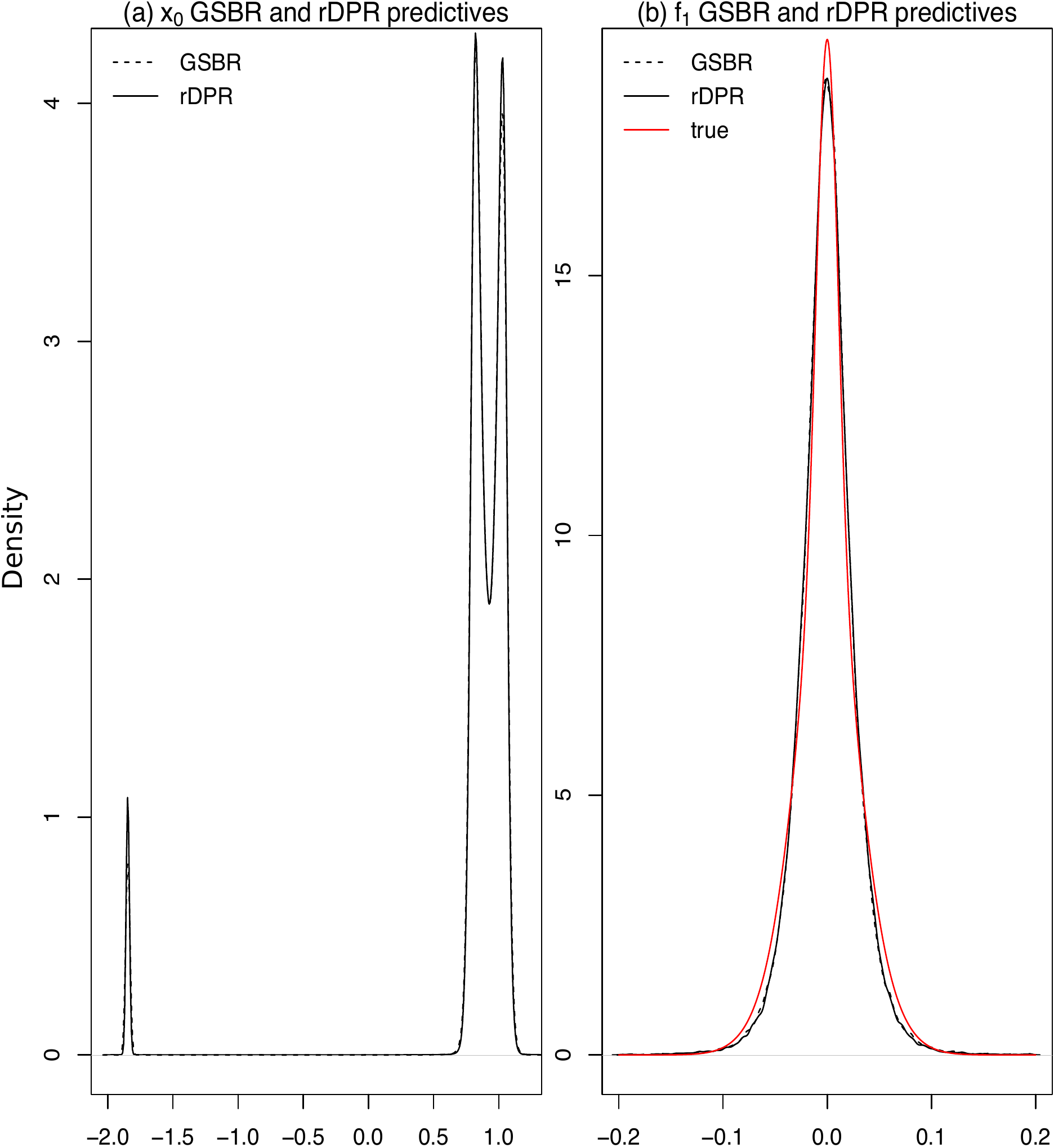}
		\caption{In Figure \ref{fig3}(a) we give superimposed the KDE's based on the posterior marginal predictive 
			samples of the initial condition variable $x_0$. In Figure \ref{fig3}(b) 
			we superimpose the GSBR and the rDPR noise density estimations together with the true dynamical 
			error density.
			\label{fig3}}
	\end{figure}

	\medskip\noindent {\bf Initial condition and dynamical noise density estimations:}  
	In Figure \ref{fig3}(a) we display kernel density estimations (KDE's) based on the predictive samples of the marginal posterior 
	for the initial condition $x_0$. The differences between the two predictives coming from the GSBR and rDPR samplers are indistinguishable.
	The three modes of the predictive density of $x_0$ are very close to the three real roots of the polynomial equation 
	$\tilde{g}(\vartheta^*,x)-\tilde{g}(\vartheta^*,1)=0$ which are the preimages of $\tilde{g}(\vartheta^*,1)$.
	Note that for $\vartheta\in(0.74, 2.97)$, it is that $\tilde{g}^{-1}(\vartheta,\tilde{g}(\vartheta,1))\in\{\rho,-1-\rho,1\}$
	with $\rho=-(1+\sqrt{4\vartheta/0.99-3})/2$.
	We refer to the three preimages of $\tilde{g}(\vartheta,1)$ by $x_{\rm L}=\rho$ (left), $x_{\rm M}=-1-\rho$ (middle) and $x_{\rm R}=1$ (right).
	In Figure \ref{fig3}(b), we give superimposed the noise predictives coming from the two models together with the true density of the noise component given in (\ref{noisedensity}). We note how the synchronized execution produces almost identical dynamical noise density estimations, 
	which are very close to the true noise density $f_1$ (solid line in red).
	
\begin{figure*}
		\centering
		\includegraphics[width = 1\textwidth]{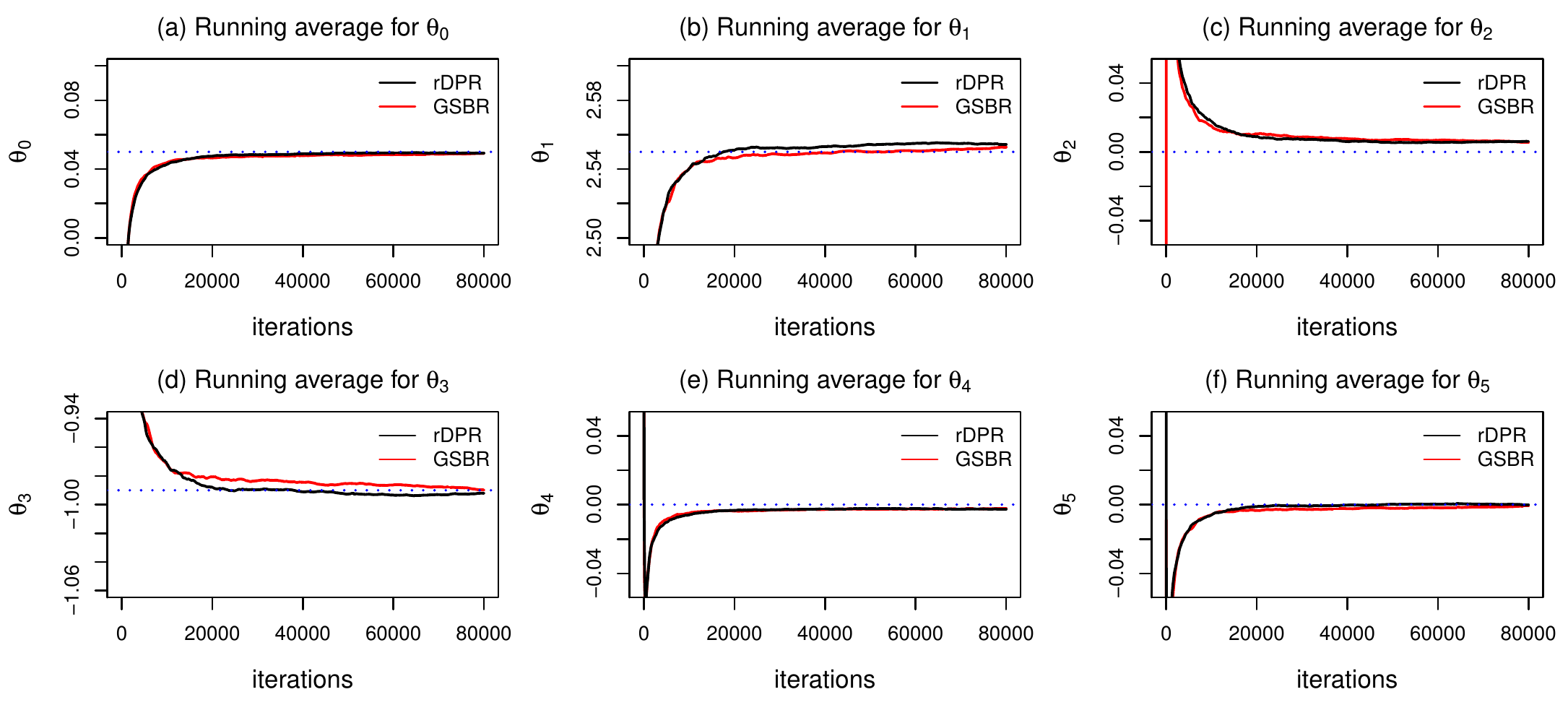}
		\caption{Chain ergodic averages for the $\theta_j$ variables based on the data set $x_{f_1}^{(100)}$, 
			under prior specification ${\cal PS}_{\rm IR}$, are superimposed in Figures \ref{fig4}(a)-(f).
			\label{fig4}}
	\end{figure*}

	In Figures \ref{fig4}(a)-(f), we plot the running ergodic averages for the $\theta_j$ variables of the first $80,000$ iterations after burn-in.
	We observe that the $\theta_j$ chains have converged after the first $10,000$ iterations, and that the chains are mixing well. 
	In Table \ref{tab1} we display the percentage absolute relative errors (PARE's) of the synchronized estimations.
	For each $j$, we have created $K=47$ approximately independent samples of size $N=10^4$, each sample separated by $s=500$ observations
	$$
	\{\theta_j^{(i_r)}:M_r+1\le i_r\le M_r+N\}\quad{\rm with}\quad M_k=(r-1)(N+s),
	$$
	for $r=1,\ldots,K$.
	Then we created $K$ realizations of the sampling mean (SM) estimator. Finally we took
	$$
	\hat{\theta}_j={1\over K}\sum_{r=1}^K{1\over N}\sum_{i=M_r+1}^{M_r+N}\theta_j^{(i)},\quad 0\le j\le 5.
	$$
	We estimate $x_0$ by the maximum a-posteriori (MAP) of the $x_0$ predictive sample, 
	by dividing the interval $[-2,2]$ into $300$ bins.
	We remark the accuracy and the closeness of the estimated $\theta$ values.
	
	\begin{table}
		\centering
		\begin{tabular}{llccccccc}
			\hline
			Model      &  $\theta_0$ & $\theta_1$ & $\theta_2$ & $\theta_3$ & $\theta_4$ & $\theta_5$ & $x_0$      \\
			\hline 
			Param.     &    1.98     &   0.37      &  0.03       &  0.58      &  0.00       &  0.04      &    $x_{\rm M}:3.87$   \\
			rDPR       &    0.81     &   0.29      &  0.01       &  0.09      &  0.04       &  0.14      &    $x_{\rm M}:0.80$   \\
			GSBR       &    0.19     &   0.27      &  0.05       &  0.04      &  0.02       &  0.18      & ~$x_{\rm R\,}:0.60$   \\                                            
			\hline\hline 
			Estim.     &$x_{201}$    & $x_{202}$   &   $x_{203}$ &  $x_{204}$ &   $x_{205}$ & GSBR-Av    & Par-Av   \\                  
			\hline                                                                                      
			{\rm SM}   &   6.43      &    7.35     &   29.70     &    5.48    &   13.68     & 12.53      &   53.49  \\ 
			{\rm MAP}  &   3.84      &   11.48     &   19.16     &    2.15    &  149.06     & 37.14      &   53.25  \\ 
			\hline
		\end{tabular}
		\caption{$(\theta, x_0)$ reconstruction PAREs ($T=0$)  under the informative prior configuration.
			\label{tab1}}
	\end{table}
	
\begin{figure*}
		\centering
		\includegraphics[width = 0.85\textwidth, height = 0.4\textwidth]{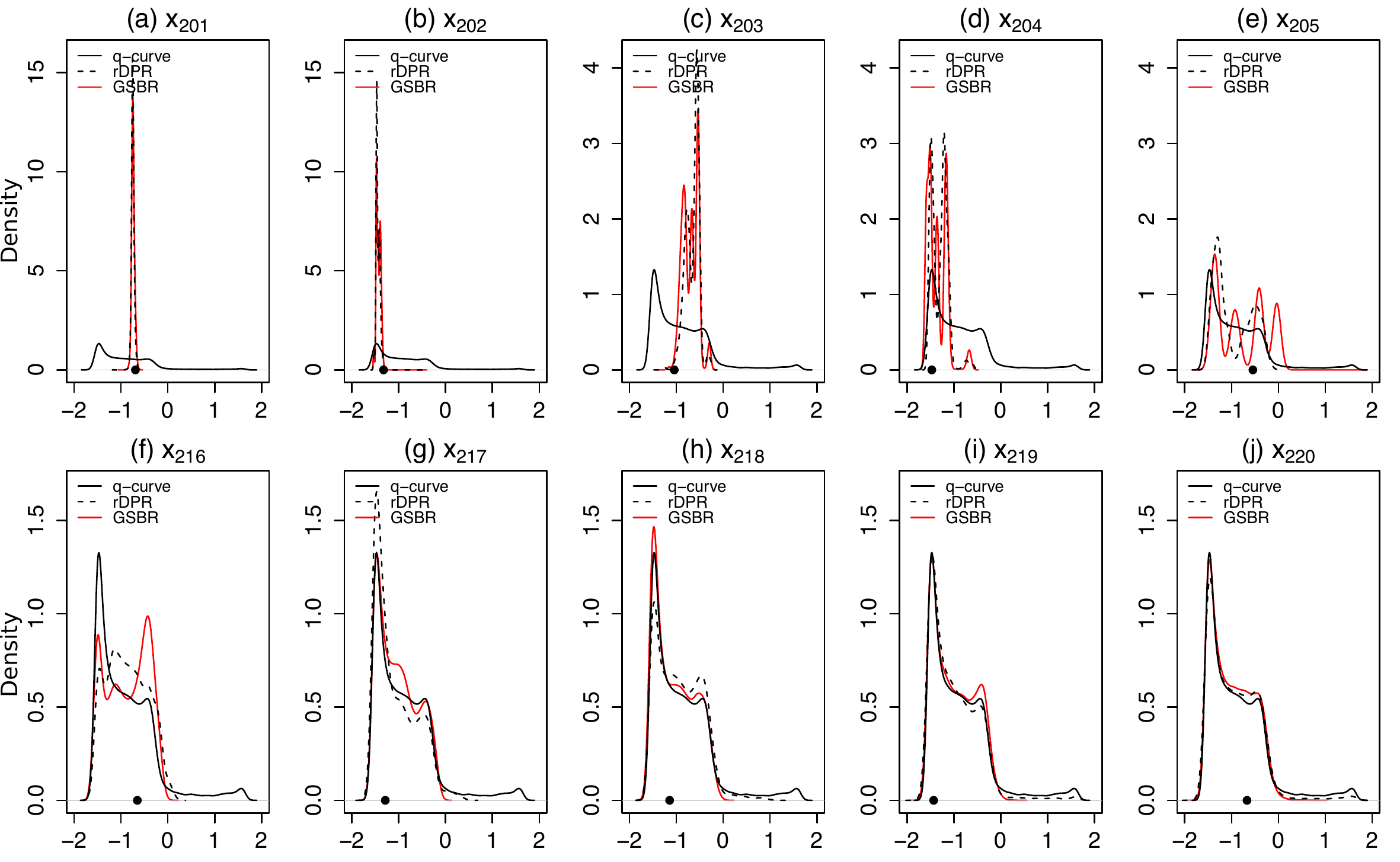}
		\caption{In Figures \ref{fig5}(a)-(j) we display superimposed the first five and the last five KDE's of the out-of-sample posterior marginal predictive 
			based on data set $X_{f_1}^{200}$ under the informative specification ${\cal PS}_{\rm IRP}$.
			Together we superimpose the KDE          
			of the $f_1$ quasi invariant density (solid black line). In all Figures, the bullet point 
			represents the corresponding true future value.\label{fig5}}            
	\end{figure*}

	\medskip\noindent {\bf Out-of-sample posterior predictive marginals and the prediction barrier:} 
	In Figures \ref{fig5}(a)-(j) we display the KDEs
	of the marginal posterior predictive samples of the variables $x_{201},\ldots,x_{205}$
	and $x_{216},\ldots,x_{220}$ coming from the GSBR (solid red line) and rDPR (dashed black line) superimposed. 
	Together, we superimpose the $f_1$ quasi-invariant measure approximation (solid black line). 
	We note how the synchronized execution produces almost identical posterior predictive marginals (PPM's).
	As the prediction horizon increases, the PPM densities are starting to resemble to the $f_1$ quasi-invariant density approximation,
	which naturally forms a prediction barrier. 
	As such, any attempt to predict beyond this time horizon will replicate the quasi-invariant measure approximation.
	From this point on, we can make only probabilistic prediction arguments for the long term behavior of the system that involve 
	the quasi-invariant measure i.e. P$(x_{n+i}\in A)=\mu_{{\tilde g},z}(A)$ for all $i\ge T$ and for all measurable subsets $A$ of $\mathbb R$.
	In table \ref{tab2}, we give the mean computational time per $10^3$ iterations relating to the synchronized execution of the rDPR and GSBR samplers under 
	prior set up ${\cal PS}_{\rm IRP}$ for a simple reconstruction ($T=0$) and prediction ($T=20$). In both cases, the GSBR sampler has the fastest execution times.
	In the last two rows of table \ref{tab1} we give the PARE's of the first five GSBR out-of-sample predictions
	using the SM and MAP estimators. The last two columns exhibit the
	mean PARE's under a GSBR and a parametric prediction.
	
	\begin{table}
		\centering
		\begin{tabular}{lccc}   
			Data set $X_{f_1}^{200}$ & & &  \\                                                                           
			\hline
			Prior spec.              & Algorithm   & $T=0$  &  $T=20$\\
			\hline
			${\cal PS}_{\rm IRP}$    &   rDPR      & 5.44    & 11.76  \\                                     
			${\cal PS}_{\rm IRP}$    &   GSBR      & 2.24    &  8.65  \\
			\hline
		\end{tabular}
		\caption{Mean execution times in seconds per $10^3$ iterations.
			\label{tab2}}
	\end{table}
	
	\subsection{Noninformative reconstruction and prediction under the $f_{2,l}$ heavy tailed dynamic noise} 
	
\begin{figure*}
			\centering
			\includegraphics[width = 1\textwidth, height = 0.5\textwidth]{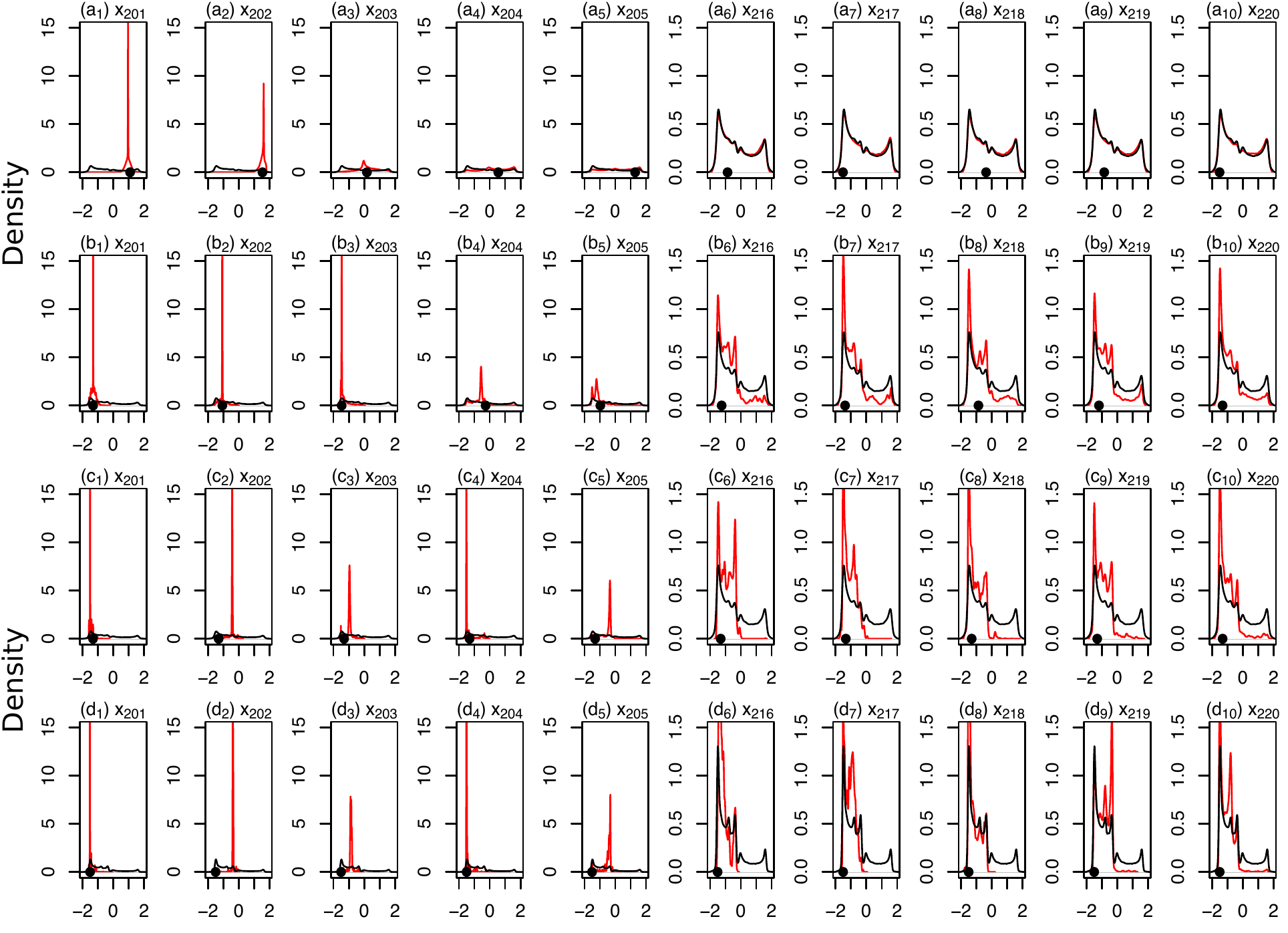}
			\caption{In Figure \ref{fig6} we display the GSBR KDE's of the PPM sample of the  
				out-of-sample variables $\{x_{201},\ldots,x_{205}\}$ and $\{x_{216},\ldots,x_{220}\}$(solid lines in red) 
				based on samples $X_{f_{2,l}}^{200}:1\le l\le 4\}$ (rows (a) to (d)) under the noninformative prior 
				specification. Together we superimpose the KDE          
				of the $f_{2,l}$ quasi-invariant densities  for $1\le l\le 4$ (solid lines in black).
				\label{fig6}}            
		\end{figure*}

	Here we simultaneously reconstruct and predict using the noninformative prior set up. 
	More specifically for $T=20$ we set $\a=\b=0.3, a=b=10^{-3}, M=M_0=10$;
	we iterated the GSBR sampler $5\times 10^5$ after a burn-in period of $10,000$.
	In Figure \ref{fig6} we display the KDE's based on the PPM samples of the  
	out-of-sample variables $\{x_{201},\ldots,x_{205}\}$ and $\{x_{216},\ldots,x_{220}\}$ (solid lines in red) 
	under data sets $X_{f_{2,l}}^{200}:1\le l\le 4\}$ (rows (a) to (d)). Together we superimpose the KDE          
	of the associated quasi-invariant densities  for $1\le l\le 4$ (solid lines in black).
	In Tables \ref{tab3} and \ref{tab4} we display a PARE summary of $(\theta, x_0)$ 
	estimations and out-of-sample prediction respectively, based on data sets $\{X_{f_{2,l}}^{200}:1\le l\le 4\}$. 
	In table \ref{tab3} we compare horizontally the PARE results coming from the GSBR and the parametric sampler;
	we notice that in all cases, the accuracy of the GSBR model is considerably higher than its parametric counterpart. 
	In all cases, the parametric algorithm predicts a quintic polynomial deterministic part.
	Also, the GSBR model precision improves as the noise model becomes more heavy tailed.
	In table \ref{tab4} when we compare the average PARE results coming from the GSBR and the parametric sampler
	(the last two columns) we notice that in all cases for both the SM and the MAP estimators, the prediction of the GSBR model
	is considerably better. We also notice, that as we move to a more heavy tailed noise model, the
	GSB prediction gradually improves and the MAP-GSBR estimator becomes more efficient. This is due to the multimodal nature of the PPM's generated by GSBR.

	\begin{table}
		\centering
		\begin{tabular}{ll|cccccc|r}
			\hline
			Noise     & Model   & $\theta_0$& $\theta_1$ & $\theta_2$ & $\theta_3$ & $\theta_4$ & $\theta_5$ &  $x_0$                \\
			\hline  
			$f_{2,1}$ & Param.  &     19.95 &     1.54   &   4.83     &  4.39      &    2.52    &    1.01    &               7.27    \\  
			& GSBR    &      0.51 &     0.01   &   0.06     &  0.02      &    0.02    &    0.00    &  $x_{\rm R\,}:0.03$   \\ 
			\hline
			$f_{2,2}$ & Param.  &      2.89 &     0.94   &   4.07     &  2.37      &    2.07    &    0.76    &               7.49    \\  
			& GSBR    &      0.54 &     0.05   &   0.06     &  0.12      &    0.03    &    0.03    &  $x_{\rm R\,}:0.03$   \\ 
			\hline
			$f_{2,3}$ & Param.  &     29.97 &     0.40   &   4.97     &  1.25      &    1.88    &    0.41    &               7.55    \\  
			& GSBR    &      0.20 &    0.04    &   0.04     &  0.13      &    0.02    &    0.04    &  $x_{\rm R\,}:0.03$   \\ 
			\hline            
			$f_{2,4}$ & Param.  &     15.57 &     1.07   &   1.33     &  3.71      &    0.43    &    1.03    &               6.40    \\  
			& GSBR    &      0.10 &    0.01    &   0.05     &  0.03      &    0.01    &    0.00    &  $x_{\rm R\,}:0.03$   \\                                     
			\hline                      
		\end{tabular}
		\caption{Simultaneous reconstruction-prediction under the noninformative prior specification. The  $(\theta, x_0)$ 
			PARE's are based on the data sets $\{X_{f_{2,l}}^{200}:1\le l\le 4\}$ for $T=20$.
			\label{tab3}  
		}
	\end{table}
	\begin{table}
		\centering
		\scalebox{0.9}{
			\begin{tabular}{ll|rrrrr||c|cc}
				\hline
				Noise      &Estim.     & $x_{201}$  & $x_{202}$  &   $x_{203}$&  $x_{204}$ &   $x_{205}$  & GSBR-Av & Par-Av   \\                  
				\hline  
				$f_{2,1}$  & {\rm SM}   &   12.50    &   0.86     &   12.57    &   44.04    &   82.11      & 30.42   &   58.72  \\ 
				& {\rm MAP}  &  12.86     &   2.10     &   77.13    &   25.89    &   39.99      & 31.59   &   69.62  \\ 
				\hline  
				$f_{2,2}$  & {\rm SM}   &   0.52     &    0.70    &    8.07    &  167.16    &   15.17      & 38.32   &   65.08  \\ 
				& {\rm MAP}  &   0.29     &    1.72    &    0.50    &  103.00    &   20.96      & 25.29   &   65.57  \\ 
				\hline 
				$f_{2,3}$  & {\rm SM}  &   0.72      &    7.99    &    0.01    &  9.74      &   49.94      & 13.68   &  233.53  \\ 
				& {\rm MAP} &  0.14       &    0.47    &    2.34    &  0.39      &    1.38      &  0.93   &  234.80  \\ 
				\hline 
				$f_{2,4}$  & {\rm SM}  &   0.24      &    1.01    &    2.95    &  3.79      &   40.25      &  9.65   &  60.69   \\ 
				& {\rm MAP} &  0.07       &    0.86    &    4.78    &  0.13      &   21.00      &  5.37   &  109.23  \\ 
				\hline                      
			\end{tabular}}
			\caption{Simultaneous reconstruction-prediction under the noninformative prior specification. The out-of-sample 
				PARE's are based on data sets $\{X_{f_{2,l}}^{200}:1\le l\le 4\}$ for $T=20$. The GSBR-Av and Par-Av
				columns are the PARE means of the first five out-of-sample estimations using the GSBR and the parametric
				Gibbs samplers respectively.
				\label{tab4}
			}
		\end{table}

		\section{Discussion}
		
		We have described a Bayesian nonparametric approach for dynamical reconstruction and prediction from observed time series data. 
		The key insight is to use the GSB process, developed by Fuentes--Garc\'ia et al. (2010), as a prior (over the space
		of densities) on the noise component. 
		The GSBR model removes a level from the hierarchy of the rDPR model as it replaces the weights of the stick breaking representation 
		of the DP with their expected values, leading to a simpler model with only one infinite dimensional parameter, the 
		locations of the atoms $(\lambda_j)$ of the random measure.
		GSB mixture dynamical 
		modeling is as accurate as DP based modeling but it gives smaller execution times, and is easier to implement.
		
		We have also shown that in a joint prediction of future values of a low dimensional noisy chaotic time series, the
		quasi-invariant set appears as a ``prediction barrier''. Also, our numerical experiments
		indicate that when the sample size of the time series is small, the 
		forecastable component analysis $\Omega$ measure can group the available sets of observations
		in terms of their complexity. A larger $\Omega$ index suggests a less informative prior set up.
		We note, that when there is strong evidence that the dynamical error has a Gaussian distribution, and the length
		of the observed time series is large, the application of nonparametric models is superfluous. Nevertheless, 
		when the size of the observed time series is very small and the dynamical errors are Gaussian, the accuracy of the simple MCMC (Param.) depends heavily on the 
		particular realization of the noise process. Then the application of the GSB based algorithm will be 
		in principle more accurate. Infinite mixtures of zero mean Gaussians, can mimic the effect of any heavy tailed symmetric noise processes, 
		of finite or infinite kurtosis to an arbitrary level of accuracy. Hence, natural directions for our future 
		research interests include:
		
		\medskip\noindent
		{\bf 1. Estimation and prediction of dynamical systems perturbed by impulsive dynamic noise:} 
		We believe that in this case the prior for the noise component should have the mixed type representation
		$$
		F(dz)=q\sum_{j=1}^\infty\pi_j\,{\cal N}(z|0,\lambda_j^{-1})dz+(1-q)\,\delta_0(dz),
		$$
		with $q\sim{\cal B}e(h_1,h_2),$ which is a random mixture of a Dirac measure concentrated at zero and a GSB mixture of
		zero mean normal kernels.
		
		\medskip\noindent
		{\bf 2. Estimation and prediction of dynamical systems under strong and persistent dynamic perturbations:} It is possible that the coexistence of a large number of active components 
		(informative prior set up) and strong and persistent dynamic noise when $T>0$, will affect the mixing properties of the 
		$\theta$ component of the Gibbs sampler; thus producing biased estimations of the $\theta_i$'s
		A possible solution to this problem could be the introduction of a $(\theta,\epsilon)$-constrained 
		Gibbs sampler together with an adaptive Gibbs scheme 
		for the out-of-sample variables. 
		More specifically, when 
		$\max_{\,0\le j\le k}\left|(\hat{\theta}_{j}^*-\hat{\theta}_{j}))/\hat{\theta}_j^*\right| > \epsilon$,
		where $\hat{\theta}^*$ is the reconstruction ($T=0$) estimation, and $\epsilon$ a small predefined constant, we propose the restriction of the $\theta$ prior specification to: 
		$$
		\theta\sim{\cal U}\left(\prod_{j=0}^k(\hat{\theta}_j^*-\epsilon_j, \hat{\theta}_j^*+
		\epsilon_j)\right),\quad \max_j\epsilon_j<\epsilon.
		$$
		When the mixing properties of the out-of-sample variables components of the Gibbs sampler are affected, 
		very long chains are needed to achieve convergence. In that case, we could resort to more sophisticated 
		Monte Carlo schemes to improve the sampling efficiency, such as hybrid Monte Carlo \cite{neal2011mcmc}, 
		or adaptive random scan sampling \cite{latuszynski2013adaptive}, in order to improve the sampling efficiency.
		
		\medskip\noindent
		{\bf 3. When the data available are contaminated with dynamical and observational noise:}
		We can extend the GSBR model to a $q$-lagged state space model, more precisely
		\begin{eqnarray}
		&  & X_i = g(\theta, X_{i-1},\ldots,X_{i-q}) + Z_i,\,\,i\ge q\nonumber\\
		&  & Y_i = h(\phi, X_i)+W_i,\nonumber
		\end{eqnarray}
		for some function $h$.
		Here we assume that noisy measurements of the output occur at all times, making the
		$X^n$ sequence unobservable. The set of observations in this case is the $Y^n$ time series, which can 
		be modeled via a GSB random measure $\P_Y$. Then the latent $X^n$ series can be modeled with a second 
		independent GSB random measure $\P_X$, such that the random variables 
		$[X_i|X_{i-1},\ldots,X_{i-p},\theta,\P_X]$ and $[Y_i|X_i,\phi,\P_Y]$ are independent.
		In this case we have to estimate the initial condition $(X_0,\ldots,X_{q-1},Y_0)$, the parameter ($\theta,\phi$), the 
		density of the noise component $(Z_i, W_i)$ as well as the hidden orbit $\{X_i:i=q,\ldots,n\}$.

		 \vspace{0.5cm}
			\noindent \textbf{{\large Acknowledgments}} \newline The second author is funded by the ``Ypatia'' scholarship program of the University of the Aegean.
				
		\section*{Supplementary material}
		In this section we provide the Appendices A,B,C and D referenced on the text.
		
		\section*{Appendix A: Sampling from nonstandard full conditionals}
		
		Here we adapt our calculations for the specific case where the deterministic part is a polynomial of degree $m$, 
		namely $g(\theta, x) = \sum_{k=0}^m\theta_k\, x^k$. 
		
		\vskip0.1in\noindent
		{\bf 1. Sampling the $\theta$-coefficients:}
		From equation (16) in the main text and for $j=1,\ldots, m$ it is that
		\begin{equation}
		\label{thetafulcond}
		f(\theta_j|\cdots)\,\propto\,{\cal I}(\theta\in\tilde{\Theta}_j)
		\exp\left\{-{1\over 2}\sum_{i=1}^n\lambda_{d_i}h_\theta(x_i,x_{i-1})\right\},
		\end{equation}
		where $\tilde{\Theta}_j$ is the $j$--th projection interval of the set $\tilde{\Theta}$. Letting
		$
		\xi_{ji}:=x_i-\sum_{\substack{k=0 \\ k\ne j}}^m\theta_k\, x_{i-1}^k,
		$
		we obtain the full conditional for $\theta_j$, which is a normal truncated over the set $\tilde{\Theta}_j$ given by
		\begin{equation}
		\label{trnormal}
		f(\theta_j|\cdots)\,\propto\,{\cal I}(\theta\in\tilde{\Theta}_j){\cal N}(\theta_j|\mu_j, \tau_j^{-1})
		\end{equation}
		with
		\begin{equation*}
		\mu_j:=\tau_j^{-1}\sum_{i=1}^n\lambda_{d_i}\xi_{ji}x_{i-1}^j,\,\,\tau_j:=\sum_{i=1}^n \lambda_{d_i}x_{i-1}^{2j}.
		\end{equation*}
		To sample from this density, a-priori we set $\theta_j\in\tilde{\Theta}_j:=(\theta_j^-,\theta_j^+)$ and we augment the $\theta_j$ full 
		conditionals by the auxiliary variables $\theta_j'$ \cite{damlen1999gibbs} such that jointly
		\begin{equation}
		\label{newjoint1}
		f(\theta_j,\theta_j'|\cdots)\propto {\cal U}(\theta_j|\theta_j^-,\theta_j^+)\,
		{\cal I}\left(\theta_j'>(\theta_j-\mu_j)^2\right)e^{-\tau_j\theta_j'/2}.
		\end{equation}
		
		Then we have the following Lemma:
		
		{\sl {\bf Lemma A.1}
			The augmentation of the full conditionals of $\theta_j$ for $j=1,\ldots,m$ with the positive random variables $\theta_j'$
			such that they jointly  satisfy (\ref{newjoint1}), leads to the following embedded Gibbs sampling scheme:

				\begin{eqnarray}
				f(\theta_j'|\theta_j,\cdots) & \propto & {\cal E}(\theta_j'|\tau_j/2)\,{\cal I}(\theta_j'>(\theta_j-\mu_j)^2) \nonumber\\
				f(\theta_j|\theta_j',\cdots) & = & \,{\cal U}(\theta_j|\alpha_j,\beta_j),\,\,
				\alpha_j:=\max\{\theta_j^-,\mu_j-\theta_j'^{1/2}\},\,\,
				\beta_j:=\min\{\theta_j^+,\mu_j+\theta_j'^{1/2}\}.\nonumber
				\end{eqnarray}

			where ${\cal E}(\theta_j'|\tau_j/2)$ denotes the exponential density with rate $\tau_j/2$.}
		\vskip0.1in\noindent
		{\bf Proof:} These are the full conditionals of the bivariate density given in equation (\ref{newjoint1}).\hfill$\square$
		
		\vskip0.1in\noindent
		{\bf Sampling the initial condition:}
		Similarly, to sample from the full conditional of $x_0$ in relation (13) given in the main text, we introduce 
		the variable $x_0'$ such that
		$$
		f(x_0,x_0'|\cdots)\propto {\cal I}(x_0\in\tilde{X})\,
		{\cal I}\left( x_0'>h_\theta(x_1,x_0)\right)e^{-\lambda_{d_1}x_0'/2}.
		$$
		Clearly, the full conditional of $x_0'$ is an exponential of rate $\lambda_{d_1}/2$, truncated over the interval $(h_\theta(x_1,x_0),\infty)$.
		The new full conditional for $x_0$ is a mixture of at most $m$ uniforms given by

			\begin{equation}\label{newfull}
			f(x_0|x_0',\cdots) \propto {\cal I}(x_0\in\tilde{X})\,{\cal I}(x_0\in{\cal R}_g),\,\,
			{\cal R}_g:=\{x:\underline{x}_{\,0}<g(\theta,x)<\overline{x}_0\},
			\end{equation}

		where $\underline{x}_{\,0}:=x_1-x_0'^{1/2}$ and $\overline{x}_0:=x_1+x_0'^{1/2}$.
		The set ${\cal R}_g$ can be represented as the union of intervals, 
		with boundaries defined by the real roots of the two polynomial equations 
		\begin{equation}
		\label{poleqs}
		\underline{q}(x_0):= g(\theta,x_0)-\underline{x}_{\,0}=0,\quad
		\overline{q}(x_0):= g(\theta,x_0)-\overline{x}_0=0.
		\end{equation}
		More specifically, we are going to show that there is $r\le m$ such that
		\begin{equation}
		\label{unionset}
		{\cal R}_g=\medcup_{i=1}^r (\rho_{2i-1},\rho_{2i}),
		\end{equation}
		with $\{\rho_1,\ldots,\rho_{2r}\}$ the ordered set of the real roots of the two polynomial equations in (\ref{poleqs}).
		In the sequel we make use of the following notation
		\begin{eqnarray}
		\{\overline{q}<0\}&:=&\{x_0\in{\mathbb R}:\overline{q}(x_0)<0\},\nonumber\\
		\{\underline{q}>0\}&:=&\{x_0\in{\mathbb R}:\underline{q}(x_0)>0\}.\nonumber
		\end{eqnarray}
		First we will consider the two even degree cases. When the leading coefficient is positive,
		the equation $\overline{q}=0$ has at least two real roots. If there are more than two real roots, their 
		number will be a multiple of two. On the other hand, when $\underline{q}=0$ has real solutions their number will be even. 
		Then for $s'\ge 1$ and $t'\ge 0$ it is that
		\begin{eqnarray}
		\label{overline1}
		\{\overline{q}<0\} & = & (\overline{\rho}_1,\overline{\rho}_2)\cup\cdots\cup(\overline{\rho}_{2s'-1},\overline{\rho}_{2s'})\\
		\label{underline1}
		\{\underline{q}>0\} & = & (-\infty,\underline{\rho}_{\,1})\cup\cdots\cup(\underline{\rho}_{\,2t'},\infty).
		\end{eqnarray}
		When $t'\ge 1$ it is that $\overline{\rho}_1<\underline{\rho}_{\,1}<\underline{\rho}_{\,2t'}<\overline{\rho}_{2s'}$. Therefore $r=2(s'+t')$ and the intersection 
		of the two sets $\{\overline{q}<0\}$ and $\{\underline{q}>0\}$ is of the form (\ref{unionset}). When the leading coefficient is negative
		the result is similar with the right hand sides of equations (\ref{overline1}) and (\ref{underline1}) interchanged.
		
		When the degree is odd and the leading coefficient is positive, both equations $\overline{q}=0$ and $\underline{q}=0$ have at least one real solution
		$\overline{\rho}_1$ and $\underline{\rho}_{\,1}$ respectively, with $\underline{\rho}_{\,1}<\overline{\rho}_1$. 
		If some of the two equations have more than one real solution, the number of the additional roots will be a multiple of two. 
		So for $s'\ge 0$ and $t'\ge 0$ it is that
		\begin{eqnarray}
		\label{overline2}
		\{\overline{q}<0\} & = & (-\infty,\overline{\rho}_1)\cup(\overline{\rho}_2,\overline{\rho}_3)\cup\cdots\cup(\overline{\rho}_{2s'},\overline{\rho}_{2s'+1})\\
		\{\underline{q}>0\} & = & (\underline{\rho}_{\,1},\underline{\rho}_{\,2})\cup\cdots\cup(\underline{\rho}_{\,2t'-1},\underline{\rho}_{\,2t'})\cup(\underline{\rho}_{\,2t'+1},\infty).\nonumber\\
		\label{underline2}
		\end{eqnarray}
		For $s'\ge 1$ and $t'\ge 1$ we have $\underline{\rho}_{\,1}<\overline{\rho}_1<\underline{\rho}_{\,2t'+1}<\overline{\rho}_{2s'+1}$, 
		and $r=2(s'+t'+1)$ which shows that the intersection 
		of the two sets $\{\overline{q}<0\}$ and $\{\underline{q}>0\}$ is of the form (\ref{unionset}).
		When the leading coefficient is negative
		the result is similar with the right hand sides of the equations (\ref{overline2}) and (\ref{underline2}) interchanged.

		So we have proved the following lemma:
		
		\vskip0.1in\noindent
		{\sl {\bf Lemma A.2}
			The augmentation of the full conditional of $x_0$ with the positive random variable $x_0'$
			leads to the following embedded Gibbs sampling scheme:
			\begin{eqnarray}
			f(x_0'|x_0,\cdots) & \propto & {\cal E}(x_0'|\lambda_{d_1}/2)\,{\cal I}( x_0'>h_\theta(x_1,x_0)) \nonumber\\
			f(x_0|x_0',\cdots) & \propto & {\cal I}(x_0\in\tilde{X})\,{\cal I}\left(x_0\in\medcup_{i=1}^r (\rho_{2i-1},\rho_{2i})\right),\nonumber
			\end{eqnarray}
			for some $r\le m$, with $\{\rho_1,\ldots,\rho_{2r}\}$ being the ordered set of the real roots of the two polynomial equations in (\ref{poleqs}).}
		
		\vskip0.1in\noindent
		{\bf 2. Sampling the first $T-1$ future observations:}
		The full conditionals $x_{n+j}$ for $1\le j\le T-1$ in relation (14) given in the main text are nonstandard densities.
		We augment the conditional of $x_{n+j}$ with the pair of variables $(x_{n+j}',x_{n+j}'')$ such that jointly

			\begin{eqnarray}
			f(x_{n+j},x_{n+j}',x_{n+j}''|\cdots) & \propto & e^{-{1\over 2}\lambda_{d_{n+j}}x_{n+j}'}\,{\cal I}(x_{n+j}'>h_\theta(x_{n+j},x_{n+j-1}))\nonumber\\
			& \times & e^{-{1\over 2}\lambda_{d_{n+j+1}}x_{n+j}''}\,{\cal I}(x_{n+j}'' > h_\theta(x_{n+j+1},x_{n+j})).\nonumber
			\end{eqnarray}

		The full conditionals of $x_{n+j}'$ and $x_{n+j}''$ are truncated exponentials with rates  $\lambda_{d_{n+j}}/2$ and  $\lambda_{d_{n+j+1}}/2$
		over the intervals $(h_\theta(x_{n+j},x_{n+j-1}),\infty)$ and $(h_\theta(x_{n+j+1},x_{n+j}),\infty)$ respectively. 
		
		The full conditional of $x_{n+j}$ is of the form (\ref{newfull}) with the set $\tilde X$ replaced by the set $(x_{n+j}^-,x_{n+j}^+)$
		with $x_{n+j}^\pm:= g(\theta, x_{n+j-1})\pm x_{n+j}'^{1/2}$, 
		and the set ${\cal R}_g$ replaced by the set $\{x:\underline{x}_{\, n+j}<g(\theta,x)<\overline{x}_{n+j}\}$
		with $\underline{x}_{\, n+j}:= x_{n+j+1}-x_{n+j}''^{1/2}$ and $\overline{x}_{n+j}:=x_{n+j+1}+x_{n+j}''^{1/2}$.

		\vskip0.1in\noindent
		{\bf 3. Sampling the geometric probability $p$ :}
		To sample from the density in relation (17) in the main text we include the pair
		of positive auxiliary random variables $p_1$ and $p_2$ such that
		$$
		f(p,p_1,p_2|\cdots)\propto p^{2n_T-\alpha-1}{\cal I}(p_1<(1-p)^{L_{n_T}}){\cal I}(p_2<e^{-\beta/p}),
		$$
		with $p\in(0,1).$
		The
		full conditionals for $p_1$ and $p_2$ are uniforms
		$$
		f(p_1\,|\cdots)={\cal U}(p_1|\,0,(1-p)^{L_{n_T}}),\quad f(p_2\,|\,\cdots)={\cal U}(p_2|\,0,e^{-\beta/p}).
		$$
		The new full conditional for $p$ becomes

			\[
			f(p\,|p_1,p_2,\cdots)\,\propto\,p^{2n_T-\alpha-1}
			\begin{cases} 
			\hfill {\cal I}\left(-{\beta\over\log p_2}<p<1-p_1^{1/L_{n_T}}\right)    \hfill & L_{n_T}\ge 0 \\
			\hfill {\cal I}\left(\max\left\{-{\beta\over\log p_2},1-p_1^{1/L_{n_T}}\right\}<p<1\right) \hfill & L_{n_T}<0. \\
			\end{cases}
			\]

		We can sample from this density using the inverse cumulative distibution function technique.
		
		We note that for a standalone application of the GSBR sampler, a beta prior distribution for $p$
		is more preferable as it leads to an implementation of a lesser complexity
		(in our paper we have chosen to assign a transformed gamma prior merely for comparison purposes).
		Letting $f(p;\a,\b)={\cal B}e(p;\alpha,\beta)\propto p^{\alpha-1}(1-p)^{\beta-1}$ and using the 
		GSB likelihood given in main text equation (11) we obtain
		$$
		f(p\,|\cdots)\,\propto\,p^{\alpha+2n_T-1}(1-p)^{\beta+\sum_{i=1}^{n_T}N_i-1},
		$$ 
		which is a beta density with shapes $\alpha+2n_T$ and $\beta+\sum_{i=1}^{n_T}N_i$.

		\section*{Appendix B: The invariant set of the 
			polynomial map $x'=\tilde{g}(\vartheta^*,x)$}
		
		For $\vartheta=\vartheta^*=2.55$ we let 
		$$
		\tilde{g}(x)\equiv\tilde{g}(\vartheta^*,x)=0.05+2.55x-0.99x^3,
		$$
		and we define $\tilde{g}^{(n)}$ to be the $n$-fold composition of $\tilde{g}$ with itself.
		We let ${\cal R}^{(2)}$ to be the set of real roots of the polynomial equation $\tilde{g}^{(2)}(x)=x$, with  $\ux=\min{\cal R}^{(2)}$,
		$\ox=\max{\cal R}^{(2)}$ and $\X=[\,\ux,\ox\,]$. We denote the complement of $\X$ by $\X'=\X'_-\cup \X'_+$,
		where $\X'_- =(-\infty,\ux\,)$ and $\X'_+ =(\ox,\infty)$. We will prove the following lemma:
		
		\vskip0.1in\noindent
		{\sl {\bf Lemma B.1} Let $\tilde g$ be the polynomial given in relation (18) of the main text, then
			for all $x\in\X'$, it is that
			$\liminf_{n\to\infty}\tilde{g}^{(n)}(x)=-\infty$ and $\limsup_{n\to\infty}\tilde{g}^{(n)}(x)=\infty$.}
		
		\smallskip\noindent
		{\bf Proof.} 
		It is not difficult to verify geometrically the following facts:
		\begin{enumerate}
			\item $\tilde{g}( \ux\,)=\ox,\,\,\tilde{g}( \ox\,)=\ux$.
			\item $\ux\,\le x\le\ox\,\Leftrightarrow\,\ux\,\le \tilde{g}(x)\le\ox$.
			\item $\tilde{g}(x)>x,\,\,\tilde{g}^{(2)}(x)<x,\,\,\forall\,x\in \X'_-$.
			\item $\tilde{g}(x)<x,\,\,\tilde{g}^{(2)}(x)>x,\,\,\forall\,x\in \X'_+$.
			\item The restrictions of $\tilde{g}$ and $\tilde{g}^{(2)}$ to $\X'$, are decreasing and increasing functions respectively.
		\end{enumerate}
		Then for all $x\in \X'_-$ we have the set of inequalities
		$$
		\tilde{g}^{(2n+1)}(x)<\tilde{g}^{(2n-1)}(x) <\cdots <\tilde{g}(x)<\ux.
		$$
		Suppose that $\lim_{n\to\infty}\tilde{g}^{(2n+1)}(x)=x^*$ then $\lim_{n\to\infty}\tilde{g}^{(2n+3)}(x)=\tilde{g}^{(2)}(x^*)=x^*$, meaning that $x^*\in{\cal R}^{(2)}$
		which is a contradiction. Therefore $\lim_{n\to\infty}\tilde{g}^{(2n+1)}(x)=-\infty$, for all $x\in \X'_-$.
		Similarly for all $x\in \X'_+$ we have the set of inequalities
		$$
		\tilde{g}^{(2n)}(x)>\tilde{g}^{(2n-2)}(x)>\cdots>\tilde{g}^{(2)}(x)>\ox,
		$$
		from which $\lim_{n\to\infty}\tilde{g}^{(2n)}(x)=\infty$, for all $x\in \X'_+$.
		\hfill$\square$
		
		\section*{Appendix C: Identification and prediction under noisy logistic observations}

		We have generated $n=200$ observations from the random logistic map via
		$$
		x_{i} = 1 - \vartheta x_{i-1}^2 + z_{i}, \,\,\, 1\le i\le 200,
		$$
		for the initial condition $x_0=1$, and the control parameter $\vartheta=\vartheta^* = 1.71.$ 
		The random dynamical error $z_i$ has been sampled independently from the noise process
		$$
		f_{2,4} = \frac{9}{10} {\cal N}(0, \sigma^2) + 
		\frac{1}{10}{\cal N}\left(0, (200\sigma)^2\right), \,\,\, \sigma = 10^{-3}.
		$$
		The time series dataset, has been generated in R under the random number generator seed $RNG_{f_{2,4}}=\{8\}$. The $f_{2,4}$ error process (see equation 20 in the main manuscript)
		produces the heaviest tail behavior as it exhibits the smaller $TF$ measure
		(see the inequalities after equation 20 in the main manuscript).
		To test the ability of the parametric model (Param.) on the identification of the correct underlying model, 
		we have modeled the associated deterministic part with a polynomial of degree $m=5$ (there
		are six $\theta$-coefficients $\{\theta_0,\ldots,\theta_5\}$).
		We ran the parametric and GSBR samplers under noninformative prior specifications for simultaneous reconstruction and prediction for $5\times 10^5$ after a burnin of $10^4$. The results are summarized in tables I and II. In Table \ref{logtab1}, we provide the percentage absolute relative errors (PARE's) for the estimation of the control parameters and in Table \ref{logtab2} we present the PARE's based on the maximum a-posteriori (MAP) estimators, for the prediction of the first 5 future unobserved observations,
		for $T=20$. In the last column of Table \ref{logtab2} we give the average PARE's obtained from the two methods. 
		
		\begin{table}[H]
			\centering
			\begin{tabular}{ll|cccccc|r}
				\hline
				Noise     & Model   & $\theta_0$& $\theta_1$ & $\theta_2$ & $\theta_3$ & $\theta_4$ & $\theta_5$ &  $x_0$\\
				\hline  
				$f_{24}$  & Param.  &  1.15 &   9.73   &   2.47   &  41.49    &    2.11    &   35.52    &   0.38  \\  
				& GSBR    &  0.04 &   0.14   &   0.18   &   0.64    &    0.32    &    0.57    &   0.12  \\                              
				\hline                      
			\end{tabular}
			\caption{Control parameter and $x_0$ PARE's based on the noisy logistic data set for $T=20$.}
			\label{logtab1}
		\end{table}

		\begin{table}[H]
			\centering
			\begin{tabular}{ll|rrrrr|c}
				\hline
				Noise      & Estim.     & $x_{201}$  & $x_{202}$  &   $x_{203}$&  $x_{204}$ &   $x_{205}$  & Average error  \\                  
				\hline  
				$f_{24}$  & Param   &   13.52    &   5.10    &   28.67    &   123.14    &   137.67      & 61.62    \\ 
				& GSBR    &    0.89    &   0.39    &    1.54    &     4.01    &     2.80      &  1.93    \\ 
				\hline                      
			\end{tabular}
			\caption{The first 5 out-of-sample PARE's based on the noisy logistic data set for $T=20$.}
			\label{logtab2}
		\end{table}
		
		Hence, it is clear that under a nongaussian noise process, the parametric Gibbs sampler cannot identify properly the true underlying model (in fact the parametric sampler predicts a quintic deterministic part). On the 
		other hand, the GSBR sampler provides us with very accurate results; for example the average GSBR PARE 
		for the $\theta$-coefficients is $0.32$ which is very small compared to the average parametric PARE which is $15.41$.

		\section*{Appendix D: Dynamical behavior of cubic map}
		
\begin{figure}[H]
				\centering
				\includegraphics[width = 0.95\textwidth]{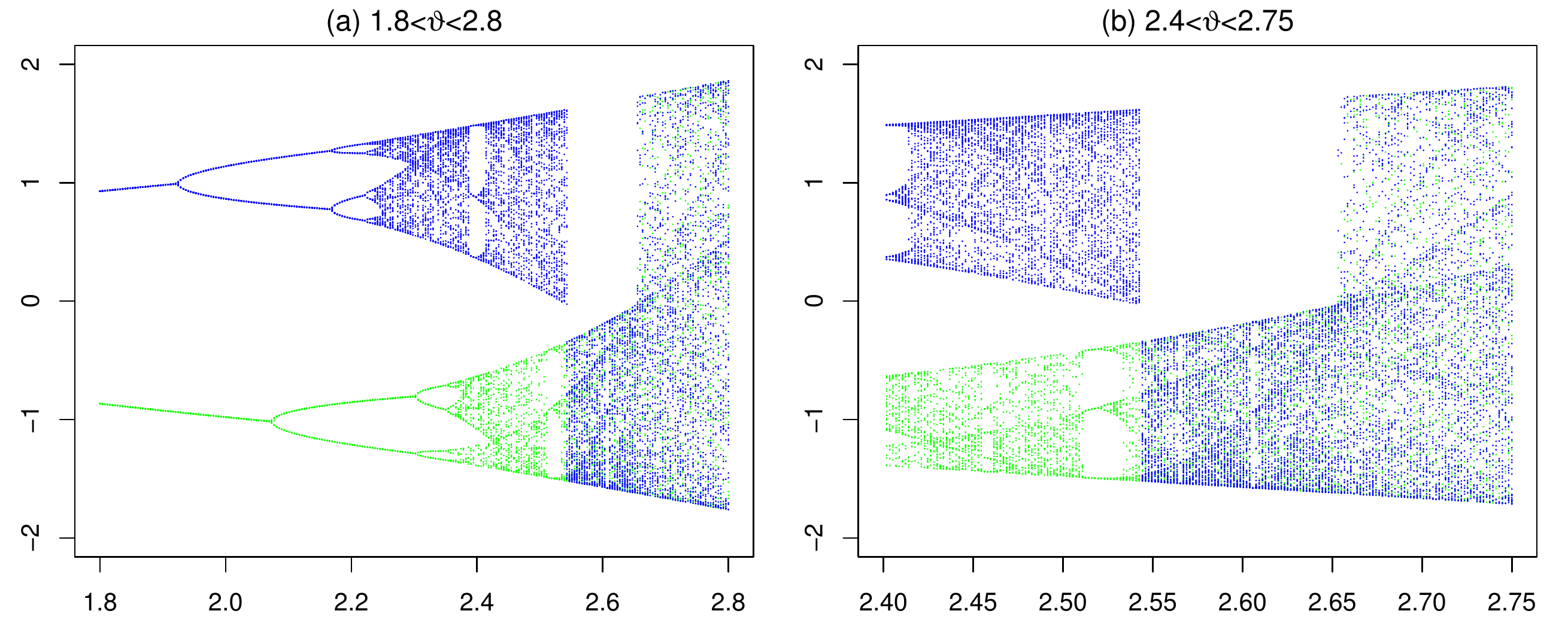}
				\caption{Bifurcation diagram of $g(\vartheta,x)=0.05+\vartheta x-0.99 x^3$.}
				\label{bifurcation}
			\end{figure}

		While quadratic polynomial maps, can exhibit for each parameter 
		value at most one stable attractor, multistability and coexistence of more than one strange attractors can be achieved under higher degree polynomial maps \cite{kraut1999preference}. In this work, we illustrate the performance of the proposed model on a cubic map with complicated dynamical behavior. In particular, we perurbed with dynamical noise the random map with deterministic part 
		\begin{equation}\label{amap2}
		g(\vartheta,x)=0.05+\vartheta x-0.99 x^3
		\end{equation}
		fixing its controlling parameter at $\vartheta=\vartheta^*=2.55$. Generally, when $\vartheta\in \Theta_{\rm bi}=[\,\underline{\vartheta}_{\,\rm bi},\overline{\vartheta}_{\rm bi}]$ with $\underline{\vartheta}_{\,\rm bi}=1.27$  and $\overline{\vartheta}_{\rm bi}=2.54$ the map becomes bistable. This means that in the phase space of the cubic map 
		we can identify for $\vartheta\in \Theta_{\rm bi}$, two mutually exclusive period-doubling cascades, together with two mutually exclusive basins of attractions. The dynamical behavior of the cubic map in (\ref{amap2}) can be depicted in the bifurcation diagram shown in Figure \ref{bifurcation}. The coexisting attracting sets are plotted in blue and green.

\begin{figure}[H]
				\centering
				\includegraphics{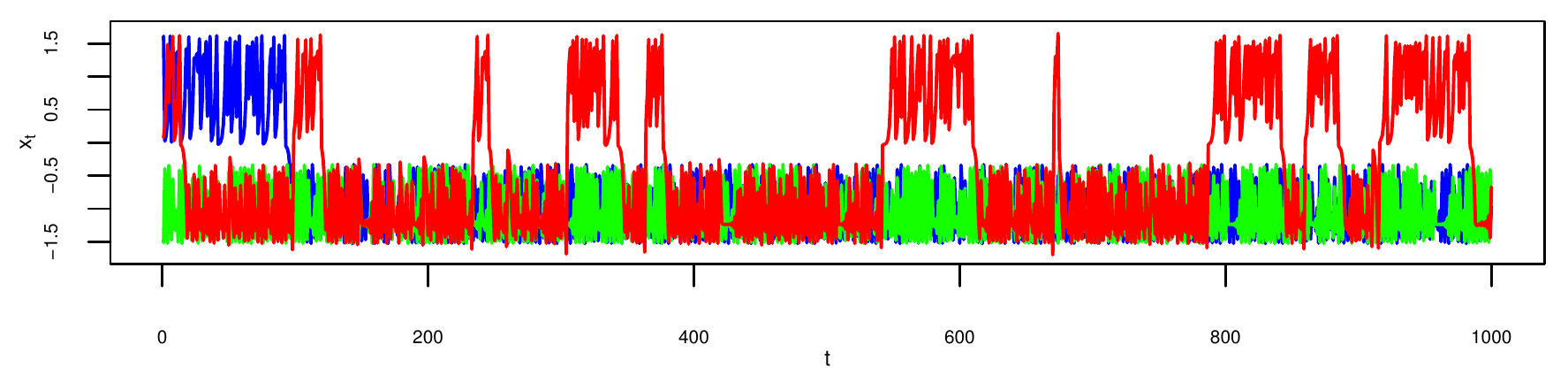}
				\caption{Orbits of $g(\vartheta,x)=0.05+\vartheta x-0.99 x^3$, with $\vartheta=2.55$. Blue and green show deterministic orbits, red shows noisy orbit.}
				\label{orbits}
			\end{figure}

		We believe that the most intricate cases in terms of system identification and prediction are emanating from the dynamically perturbed time series when $\vartheta\in [2.54,2.65]$. This is the case for $\vartheta=2.55$, the value of the control parameter we have used in our numerical experiments. For this specific value of $\vartheta$ we have the coexistence of a repelling strange set ${\cal O}_{unst,\infty}^+$ and an attracting strange set
		${\cal O}_{st,\infty}^-$. Letting $g(\vartheta,\cdot)\equiv g(\cdot)$, one has that 
		$$
		{\cal O}_{\rm unst, \infty}^+\subset\bigcup_{r\ge 1}g^{(-r)}({\cal O}_{\rm st, \infty}^-).
		$$	          
		and all orbits will be eventually attracted by the ``lower'' part ${\cal O}_{st,\infty}^-$. 
		Nevertheless when the $f_{2,4}$ dynamical noise is present, the random orbits (in red) are able to visit the vicinity of the repelling set ${\cal O}_{\rm unst, \infty}^+$, ad infinitum, as we show in Figure \ref{orbits}.

\bibliographystyle{plain}
\providecommand{\noopsort}[1]{}\providecommand{\singleletter}[1]{#1}%

\end{document}